\documentclass[aps,prd,floatfix,superscriptaddress,amsmath,amssymb,twocolumn,nofootinbib]{revtex4-1}
\pdfoutput=1
\usepackage{comment}
\usepackage{mathtools}
\usepackage{xcolor}
\usepackage{graphicx}
\usepackage[colorlinks=true,citecolor=blue,linkcolor=blue,breaklinks=true]{hyperref}

\begin{document}
\count\footins = 1000
\title{Mapping the evolution of supernova-neutrino-boosted dark matter within the Milky Way}

\author{Yen-Hsun Lin}
\email{yenhsun@phys.ncku.edu.tw}
\affiliation{Institute of Physics, Academia Sinica, Taipei 115, Taiwan}

\author{Meng-Ru Wu}
\email{mwu@as.edu.tw}
\affiliation{Institute of Physics, Academia Sinica, Taipei 115, Taiwan}
\affiliation{Institute of Astronomy and Astrophysics, Academia Sinica, Taipei 106, Taiwan}
\affiliation{Physics Division, National Center for Theoretical Sciences, Taipei 106, Taiwan}

\begin{abstract}

Supernova-neutrino-boosted dark matter (SN$\nu$~BDM) has emerged as a promising portal for probing sub-GeV dark matter. In this work, we investigate the behavior of BDM signatures originating from core-collapse supernovae within the Milky Way (MW) over the past one hundred thousand years, examining both their temporal evolution and present-day spatial distributions. We show that 
while the MW BDM signature is approximately diffuse in the nonrelativistic regime, it exhibits significant temporal variation and spatial localization when the BDM is relativistic. 
Importantly, we compare these local MW signatures with the previously proposed diffuse SN$\nu$~BDM (DBDM), which arises from the accumulated flux of all past supernovae in the Universe [Y.-H~Lin and M.-R.~Wu, \href{https://doi.org/10.1103/PhysRevLett.133.111004}{Phys.~Rev.~Lett. {\bf 133}, 111004 (2005)}]. In the nonrelativistic limit, DBDM consistently dominates over the local diffuse MW BDM signature. Only when the MW BDM becomes ultrarelativistic and transitions into a transient, highly-localized
signal can it potentially surpass the DBDM background. This work thus reinforces the importance of DBDM for SN$\nu$~BDM searches until the next galactic SN offers new opportunities.

\end{abstract}
\maketitle

\section{Introduction}
Although strong evidence suggests that dark matter (DM) is both abundant and stable in our Universe, its fundamental nature, whether particlelike or wavelike, remains elusive.
Over the past decades, numerous DM candidates have been proposed, ranging from ultralight axionlike particles \cite{Choi:2020rgn} to weakly interacting massive particles (WIMPs) \cite{Arcadi:2017kky}.
These theoretical developments have been accompanied by a variety of experimental search strategies, including direct detection   \cite{LUX:2016ggv,LUX:2017ree,SuperCDMS:2018mne,XENON:2018voc,XENON:2019gfn,XENON:2019zpr,SENSEI:2019ibb}, indirect searches  \cite{AMS:2015azc,IceCube:2016dgk,Fermi-LAT:2017opo,IceCube:2021xzo,IceCube:2023ies,John:2023knt,IceCube:2024yaw}, and collider-based approaches \cite{Marsicano:2018glj,Boveia:2018yeb}.
While stringent constraints have been placed on WIMPs with masses $m_\chi$ above the GeV scale, the sub-GeV regime remains 
challenging
due to the associated low-recoil signals. 
Despite these difficulties, an increasing number of novel detection schemes have emerged, leveraging new materials such as superconductors, quantum devices, and other mechanisms \cite{Hochberg:2022apz,Essig:2011nj, Graham:2012su, Essig:2015cda, Hochberg:2015pha, Hochberg:2015fth, Hochberg:2019cyy, Hochberg:2021ymx, Hochberg:2021yud, Derenzo:2016fse, Schutz:2016tid, Hochberg:2016ntt, Essig:2016crl, Hochberg:2017wce, Cavoto:2017otc, Griffin:2018bjn, Sanchez-Martinez:2019bac, Essig:2019xkx, Emken:2019tni, Hochberg:2022apz, Kurinsky:2019pgb, Blanco:2019lrf, Griffin:2020lgd, Blanco:2021hlm, Baxter:2019pnz,Kurinsky:2020dpb,Geilhufe:2019ndy,Emken:2017erx,Emken:2017qmp,Catena:2019gfa,Liang:2024lkk,Radick:2020qip,Gelmini:2020xir,Trickle:2020oki,Knapen:2021run,Cook:2024cgm, Simchony:2024kcn, Essig:2022dfa, Das:2022srn, Das:2024jdz, Griffin:2024cew, QROCODILE:2024zmg, Marocco:2025eqw}.
Innovative uses of neutrino detectors, exploiting dark counts \cite{Leane:2025efj} and solar reflection \cite{An:2017ojc} are also
proposed.

Recently, scenarios where DM is upscattered by high-energy cosmic particles have attracted growing interest.
Such boosted dark matter (BDM) offers a promising avenue to probe sub-GeV DM in large underground neutrino and DM experiments 
\cite{Bringmann:2018cvk,Ema:2018bih,Cappiello:2019qsw,Dent:2019krz,Wang:2019jtk,Zhang:2020nis,Guo:2020drq,Ge:2020yuf,Cao:2020bwd,Jho:2020sku,Cho:2020mnc,Lei:2020mii,Guo:2020oum,Xia:2020apm,Dent:2020syp,Ema:2020ulo,Flambaum:2020xxo,Jho:2021rmn,Das:2021lcr,Bell:2021xff,Chao:2021orr,Ghosh:2021vkt,Feng:2021hyz,Wang:2021nbf,Xia:2021vbz,Wang:2021jic,Lin:2022dbl,Super-Kamiokande:2022ncz,PandaX-II:2021kai,CDEX:2022fig,Granelli:2022ysi,Cline:2022qld,Xia:2022tid,Cappiello:2022exa,Carenza:2022som,Lin:2023nsm,Wang:2023wrx,COSINE-100:2023tcq,Lin:2024vzy,Xia:2024ryt,Guha:2024mjr,Kim:2024ltz,Ghosh:2024dqw,Sun:2025gyj} such as Super-Kamiokande \cite{Super-Kamiokande:2016yck}, Hyper-Kamiokande \cite{Hyper-Kamiokande:2018ofw}, DUNE \cite{DUNE:2020ypp}, and JUNO \cite{JUNO:2021vlw}, due to the sufficient energy deposition and flux required to achieve plausible detection significance within a reasonable exposure time.

Among various BDM scenarios, supernova-neutrino-boosted dark matter (SN$\nu$~BDM) \cite{Lin:2022dbl,Lin:2023nsm,Lin:2024vzy,Sun:2025gyj} offers a unique way for possible extraction of $m_\chi$ by utilizing the temporal profile of BDM flux that arrived after the detection of SN$\nu$ when the next galactic core-collapse SN (CCSN) takes places~\cite{Lin:2022dbl,Lin:2023nsm}.
Besides individual CCSN, the accumulated SN$\nu$~BDM flux from all past supernovae (SNe) throughout cosmic history contributes to a diffuse BDM (DBDM) component \cite{Lin:2024vzy}. 
This component is analogous to the well-known diffuse SN$\nu$ background 
\cite{Horiuchi:2008jz,Beacom:2010kk}, and can be readily explored by current and upcoming neutrino and DM experiments before the next galactic CCSN occurs. 

While the above works mainly focus on the SN$\nu$~BDM whose kinetic energy $T_\chi\gtrsim\mathcal{O}$(MeV), comparable to the energy scale of SN$\nu$, Ref.~\cite{Sun:2025gyj} recently discussed the SN$\nu$~BDM flux from past CCSNe in the Milky Way (MW) for $T_\chi\lesssim\mathcal{O}$(MeV) and its relevance to semiconductor-based DM experiments.  
One key assumption taken in Ref.~\cite{Sun:2025gyj} is to approximate this component as completely time-independent  
when the typical SN$\nu$~BDM duration ($t_{\rm van}$) from a single CCSN is longer than the average time interval between two successive CCSNe in MW, $\tau_{\rm CCSN}$, by taking a time-averaged SN$\nu$ spectrum across the MW as the boosting source. 
In this work, we aim to fully address this assumption by examining the full temporal and spatial dependence of SN$\nu$~BDM in the MW across a wide range of $m_\chi/T_\chi$. 
By performing Monte Carlo (MC) simulation to generate the locations and ages of $\sim 1,600$ CCSNe in the MW over the past one hundred thousand years, 
we compute the fine-resolved spatial distribution at present time and the temporal evolution of SN$\nu$~BDM flux over a span of $\sim 1,000$~yrs by summing over the contribution from all CCSNe. 
Our results quantify that the all-sky, temporal flux of the MW SN$\nu$~BDM varies by a factor of $\lesssim 5$ when $m_\chi/T_\chi \gtrsim \mathcal{O}(0.1)$, which corresponds to $t_{\rm van}\gtrsim \mathcal{O}(800)~{\rm yrs}\gtrsim 13\tau_{\rm CCSN}$.\footnote{We note that a similar requirement has been recently suggested for estimating the amount of ``diffuse SN$\nu$ component'' due to SN$\nu$ deflected by heavy DM with $m_\chi\gtrsim \mathcal{O}(1)$~GeV in MW~\cite{Chauhan:2025hoz}.} 
For $t_{\rm van}\simeq \tau_{\rm CCSN}$, the all-sky SN$\nu$~BDM flux still varies substantially over time and cannot be approximated by the averaged flux. 
Similarly, the contributions from individual CCSNe to the present-day angular distribution of SN$\nu$~BDM largely overlap only for $m_\chi/T_\chi \gtrsim \mathcal{O}(0.1)$ and can be considered as diffuse. 
For $t_{\rm van}\lesssim \tau_{\rm CCSN}$, SN$\nu$~BDM from different CCSNe can, in principle, be separated if good angular resolution in experiments can be provided. 
In addition, we find that the diffuse SN$\nu$~BDM flux from past CCSNe in the MW computed here is generally smaller than the DBDM flux
%at the same range 
for $m_\chi/T_\chi \gtrsim \mathcal{O}(0.01)$ by a factor of a few, suggesting that DBDM remains the dominant component for diffuse SN$\nu$~BDM searches. 
 
The structure of this paper is as follows.
In Sec.~\ref{sec:SNe_dist}, we describe the MC simulation of CCSN distributions in the MW over the past 100,000 years.
A brief overview of SN$\nu$~BDM physics are provided in Sec.~\ref{sec:BDM_review} for completeness. 
Numerical results illustrating the temporal and spatial distributions of BDM
are presented in Sec.~\ref{sec:numerical_results} and are compared to the DBDM flux.
In addition, given that hundreds of supernova remnants (SNRs) have been observed, we also performed a parallel analysis using this realistic dataset in Sec.~\ref{sec:histroical_SNRs} and discuss the implication.
We summarize our findings in Sec.~\ref{sec:summary}.

All calculations presented in this study were performed using our Python package \texttt{snorer} \cite{snorer2024}.
A comprehensive user manual is also provided \cite{docs}.

\section{Sampling supernova distribution in Milky Way}\label{sec:SNe_dist}

\begin{figure}
\begin{centering}
\includegraphics[width=0.6\columnwidth]{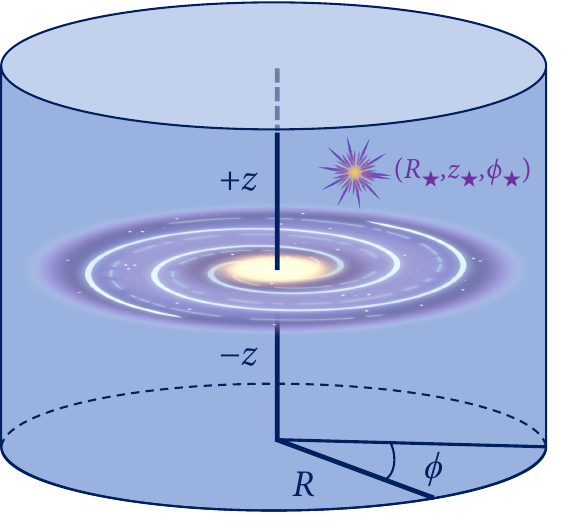}
\end{centering}
\caption{\label{fig:cylind_coord}
Schematic plot for the coordinate system used to generate the mock SN dataset in this work.
The GC is located at the origin of the cylindrical coordinate system, where $(R,z,\phi)$ correspond to the radial distance, vertical height, and azimuthal angle, respectively. The coordinates $(R_\star,z_\star,\phi_\star)$ indicate the position of a SN.
}
\end{figure}

To compute the SN$\nu$~BDM flux contribution from past CCSNe 
in the MW,
we generate a set of data that records the time and location of each CCSN over the past $10^5$~yrs with the following procedure.
We take the CCSN rate  
$R_{\rm CCSN} = 1/\tau_{\rm CCSN} \approx 1.63$ per century~\cite{Rozwadowska:2020nab} and assume a uniform probability distribution function in time. 
For the location, we assume a 
spatial distribution following the stellar density profile of the MW.\footnote{Other measures that are often used to model the spatial distribution of MW's recent CCSNe include the known distribution of dust~\cite{Adams:2013ana} as well as that of molecular hydrogen gas which may represent a better tracer of recent star formation history~\cite{Kennicutt:2012ea,Galbany:2017hzk}. 
In Appendix~\ref{app:H2_density}, we show that our results are insensitive to the choice of the underlying spatial distribution.}
We take the cylindrically symmetric MW stellar density profiles of the bulge and disc components, $\rho_b$ and $\rho_d$, respectively,   
from Ref.~\cite{McMillan:2016jtx}, 
\begin{equation}\label{eq:rho_budge}
    \rho_b(R,z)=\frac{\rho_{0,b}}{[1+r^\prime(R,z)/r_0]^\alpha}\exp\left[-\frac{r^{\prime 2}(R,z)}{r_{\rm cut}^2}\right]
\end{equation}
and
\begin{equation}\label{eq:rho_disc}
    \rho_d(R,z)=\frac{\Sigma_0}{2z_d}\exp\left(-\frac{|z|}{z_d}-\frac{R}{R_d}\right). 
\end{equation} 
Here, $R$ and $z$ denote the cylindrical radial and vertical coordinates relative to the Galactic Center (GC) (see Fig.~\ref{fig:cylind_coord}).
For the bulge, $r^\prime = \sqrt{R^2+(z/q)^2}$, with parameters $\alpha = 1.8$, $r_0 = 0.075$ kpc, $r_{\rm cut} = 2.1$ kpc, $q = 0.5$, and $\rho_{0,b} = 9.93\times 10^{10}\,M_\odot\,{\rm kpc}^{-3}$.
For the disc, we further distinguish between thin and thick components, $\rho_{d}^{\rm thin}$ and $\rho_{d}^{\rm thick}$, using parameters listed in Table~\ref{tab:stellar_disc} in Eq.~\eqref{eq:rho_disc}, respectively.

\begin{table}
\begin{centering}
\begin{tabular}{|c|c|c|c|}
\hline 
 & $\Sigma_{0}$ ($M_{\odot}\,{\rm kpc}^{-2}$) & $R_{d}$ (kpc) & $z_{d}$ (kpc)\tabularnewline
\hline 
\hline 
thin & $8.96\times10^{8}$ & 2.5 & 0.3\tabularnewline
\hline 
thick & $1.83\times10^{8}$ & 3.02 & 0.9\tabularnewline
\hline 
\end{tabular}
\par\end{centering}
\caption{Parameter values describing the density distribution of the stellar disc components in MW.\label{tab:stellar_disc}}
\end{table}

The normalized spatial probability distribution function for SNe, $p_{\rm SN}$, is given by
\begin{equation}\label{eq:p_SN}
   p_{\rm SN} (R,z) = \mathcal{N} R [\rho_b(R,z) + \rho_{d}^{\rm thin}(R,z) + \rho_{d}^{\rm thick}(R,z)], 
\end{equation}
where the normalization constant $\mathcal{N}$ satisfies  
\begin{equation}\label{eq:prob_conservation}
    2\pi \int_0^{R_{\rm max}} dR\, \int_{-z_{\rm max}}^{z_{\rm max}} dz\, p_{\rm SN}(R,z) = 1.  
\end{equation}
Since $\rho_b$ and $\rho_d$ rapidly fall off exponentially beyond $(R, z) \sim (40, 5)$ kpc, we take $(R_{\rm max}, z_{\rm max}) = (50, 10)$ kpc in Eq.~\eqref{eq:prob_conservation}. 

Given the above temporal and spatial distributions,  
we use the Python MC sampling package \texttt{emcee} \cite{emcee} to 
obtain the times and locations of 1,667 CCSNe in our dataset spanning over a time duration of $0\leq t\leq 10^5$~yrs, where $t=10^5$~yr denotes the present time. 
As we focus exclusively on CCSNe, we refer to these events as SNe throughout the rest of the paper for simplicity.

\begin{figure}
\begin{centering}
\includegraphics[width=0.9\columnwidth]{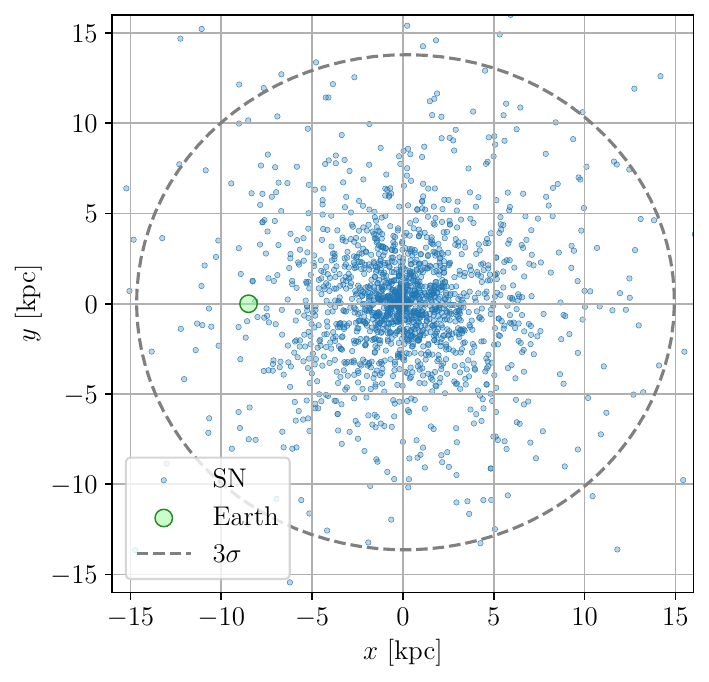}
\end{centering}
\caption{\label{fig:sne_scattering}
Locations of SNe (blue dots) over the past 100,000 years across the MW projected on the $x$-$y$ plane. The Earth, shown as a green circle, is located at $(x, y, z) = (-R_e, 0, 0)$ in Cartesian coordinates. 
The gray dashed line encloses the $3\sigma$ range of SN occurrence probability. A total of 1,667 SNe are included in this dataset.
}
\end{figure}

\begin{figure}
\begin{centering}
\includegraphics[width=0.9\columnwidth]{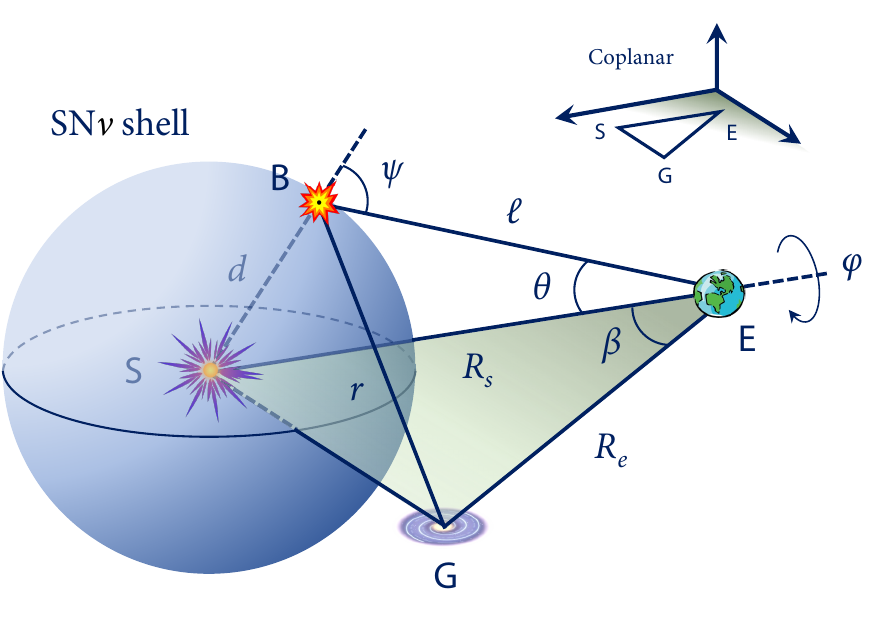}
\end{centering}
\caption{\label{fig:scheme}
Schematic plot showing the geometry and coordinates used to compute the SN$\nu$ BDM. SN ($\mathsf{S}$), GC ($\mathsf{G}$), and Earth ($\mathsf{E}$) lie on the same plane. Halo DM is upscattered by SN$\nu$ at location $\mathsf{B}$ on the outward-propagating SN$\nu$ shell.
}
\end{figure}

Fig.~\ref{fig:sne_scattering} shows the spatial locations (blue dots) of these 1,667 SNe 
in our sample projected on the $x$-$y$ plane, where $x=R\cos\phi$ and $y=R\sin\phi$. Also shown is the location of the Earth, chosen to be on the $-x$ axis with coordinate $(R,z,\phi)=(R_e,0,-\pi)$ or $(x,y,z)=(-R_e,0,0)$. 
$R_e = 8.5$~kpc is the distance between Earth and the GC.  
The gray dashed contour in Fig.~\ref{fig:sne_scattering} encloses the $3\sigma$ probability range.
Although the projected plot does not show the distribution in $z$, 
we take that into account in our calculation.
For the age of each SN, it is given by 
\begin{equation}
    t_{\rm age} = 10^5~{\rm yrs} - t_{\rm SN},
\end{equation}
where $t_{\rm SN}$ is the occurring time of the SN.

\section{SN$\nu$ BDM in Milky Way}\label{sec:BDM_review}

In this section, we review the basic framework of SN$\nu$~BDM~\cite{Lin:2022dbl,Lin:2023nsm,Lin:2024vzy} and extend the discussion on the signal duration of SN$\nu$~BDM from a single SN, 
$t_{\rm van}$, to cover the nonrelativistic limit with $m_\chi/T_\chi\gg 1$. 
Throughout this work, we adopt the assumption of an energy-independent DM-$\nu$ cross section for demonstration, noting that the same framework can be applied to any generic and energy-dependent 
particle models predicting DM-visible sector interactions \cite{He:1991qd,Fox:2008kb,Falkowski:2009yz,Lindner:2010rr,Davoudiasl:2012ag,Chang:2014tea,GonzalezMacias:2015rxl,Battaglieri:2017aum,Chang:2018rso,Foldenauer:2018zrz,Blennow:2019fhy,Escudero:2019gzq,Croon:2020lrf,Lin:2021hen,Workman:2022,Nguyen:2025ygc} as done in Ref.~\cite{Lin:2023nsm}.

To compute the SN$\nu$~BDM flux at Earth, we take a different coordinate system shown in Fig.~\ref{fig:scheme}, where Earth ($\mathsf{E}$) is located at the origin. 
We use $R_s$ ($R_e$) to denote the distance between SN at $\mathsf{S}$ (GC at $\mathsf{G}$) and Earth at $\mathsf{E}$. 
DM can be boosted at any location $\mathsf{B}$ on the expanding spherical shell of SN$\nu$ whose radius with respect to the SN location is $d$. 
The resulting BDM that arrives at Earth needs to travel a distance $\ell$  
along the direction defined by the scattering angle $\psi$. 
The angle $\theta$ denotes the viewing angle at Earth relative to the direction of the SN. 
It is important to note that $n_\chi(r)$, which is assumed to follow the NFW profile~\cite{Navarro:1995iw,Navarro:1996gj},\footnote{We do not consider the potential presence of the DM spike around the central supermassive black hole of the MW \cite{Gondolo:1999ef,Ullio:2001fb} in the main text. 
The impact of DM spike on the resulting BDM flux in the diffuse limit is discussed in Appendix~\ref{app:DM_spike}.
}
is not azimuthally symmetric in $\varphi$ unless the SN lies along the $\overline{\mathsf{GE}}$ axis so that the angle $\beta=0$. See Appendix~B in Ref.~\cite{Lin:2023nsm} for detailed discussions.
With the geometry specified above, the associated temporal profile and the angular distribution (viewed from Earth) of SN$\nu$ BDM flux 
can be evaluated unambiguously.

\begin{figure*}
\begin{centering}
\includegraphics[width=0.9\columnwidth]{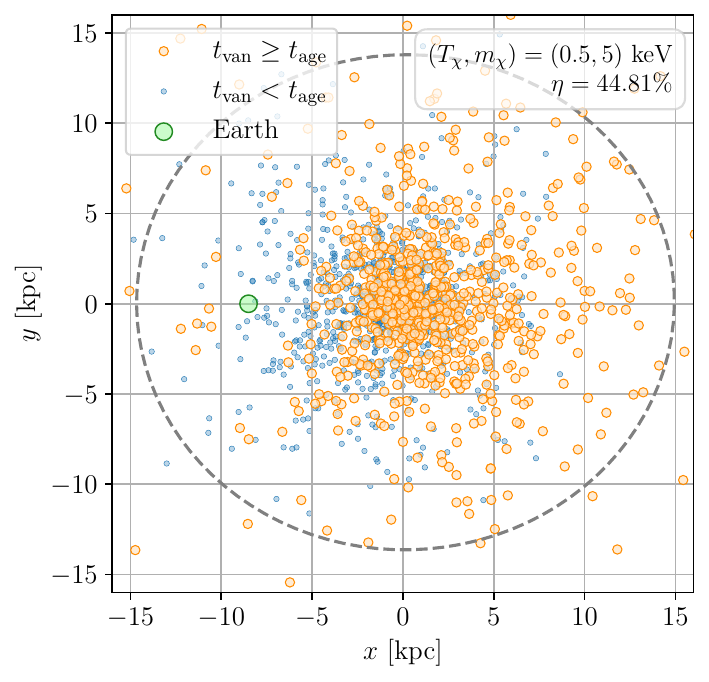}
\includegraphics[width=0.9\columnwidth]{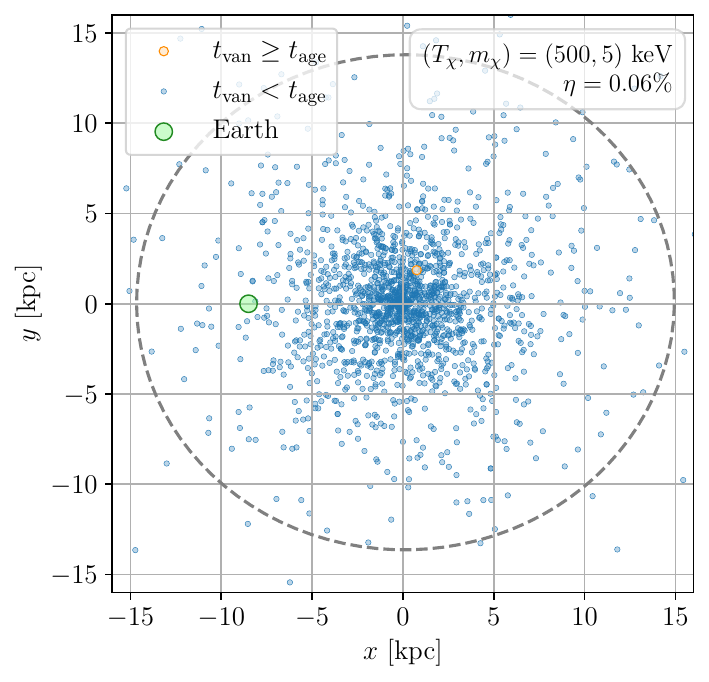}
\end{centering}
\caption{\label{fig:SNe_with_non_vanishing_flux}
Spatial distribution of SNe that contribute (orange circles) and do not contribute (blue circles) 
to the present-day BDM flux at Earth.  
$T_\chi = 0.5$~keV and 5~keV for the left and right panels. DM mass is taken to be $m_\chi=5$ keV for both panels.  
$\eta$ denotes the fraction of SNe contributing nonzero BDM flux from the whole dateset.}
\end{figure*}

\subsection{BDM flux at Earth}
The BDM emissivity at $\mathsf{B}$ is given by~\cite{Lin:2023nsm}
\begin{equation}\label{eq:jx} 
j_{\chi}(r,d,T_{\chi},\psi) =
cn_{\chi}(r)\left(\frac{1}{2\pi}\frac{d\sigma_{\chi\nu}}{d\cos\psi}\right)
\frac{dn_{\nu}}{dE_{\nu}}
\left(\frac{dE_{\nu}}{dT_{\chi}}\frac{v_\chi}{c}\right),
\end{equation}
where $E_\nu$ is the neutrino energy, $c$ is the speed of light, and $T_\chi$ is the kinetic energy transferred to DM, expressed as
\begin{equation}\label{eq:Tx}
T_{\chi}=\frac{E_{\nu}^2}{E_{\nu}+m_{\chi}/2}\left(\frac{1+\cos\theta_{c}}{2}\right),
\end{equation}
with $\theta_{c}=2\tan^{-1}(\gamma\tan\psi)$ being the scattering angle in the center-of-mass frame of scattering and $\gamma=(E_{\nu}+m_{\chi})/\sqrt{m_{\chi}(2E_{\nu}+m_{\chi})}$.  
The BDM velocity  $v_\chi/c = \sqrt{T_\chi(2m_\chi+T_\chi)}/(m_\chi+T_\chi)$.  
The differential DM-$\nu$ cross section (assumed to be uniform in the center-of-mass frame\footnote{For completeness, we demonstrate the impact of taking an energy-dependent $\sigma_{\chi\nu}$ based on the $L_\mu-L_\tau$ model \cite{Chang:2018rso,Croon:2020lrf,Escudero:2019gzq,Foldenauer:2018zrz} on the diffuse case in Appendix~\ref{app:energy_dependent_sigxv}.}) 
takes the form  
\begin{equation}\label{eq:diff_sigxv}
\frac{1}{2\pi}\frac{d\sigma_{\chi\nu}}{d\cos\psi}=\sigma_0 g_\chi(\psi),
\end{equation}
where
\begin{equation}\label{eq:gx}
    g_\chi(\psi)= \frac{\gamma^2\sec^{3}\psi}{\pi(1+\gamma^{2}\tan^{2}\psi)^{2}}
\end{equation}
is the lab-frame angular distribution at $\mathsf{B}$ for scattering angle $\psi \in [0,\pi/2]$ \cite{Lin:2022dbl}.  
In the following discussion, we fix $\sigma_0 = 10^{-39}~{\rm cm}^2$ for demonstration, which satisfies the the energy-independent $\sigma_{\chi \nu}$ constraint derived from MW satellites and subhalos for $m_\chi$ around the keV region \cite{Akita:2023yga}.

For the SN$\nu$ number density at $\mathsf{B}$, we take 
\begin{equation} \label{eq:dnv/dEv}
\frac{dn_{\nu}}{dE_{\nu}} =\sum_{i}\frac{L_{\nu_{i}}}{4\pi d^2\langle E_{\nu_{i}}\rangle c}E_{\nu}^2 f_{\nu_{i}}(E_{\nu}),
\end{equation}
where $L_{\nu_{i}} = L_{\nu,{\rm tot}}/6 \approx 5\times 10^{51}~{\rm erg\,s^{-1}}$ is the luminosity of each flavor ($\nu_e$, $\nu_\mu$, $\nu_\tau$, and their antiparticles).   
The average energies are taken as $\langle E_{\nu_e}\rangle = 11$ MeV, $\langle E_{\bar\nu_e}\rangle = 16$ MeV, and $\langle E_{\nu_x}\rangle = 25$ MeV, with $\nu_x \in \{ \nu_\mu, \nu_\tau, \bar\nu_\mu, \bar\nu_\tau \}$.  
The function $f_{\nu_i}$ is the Fermi-Dirac distribution with pinch parameter $\eta_{\nu_i} \equiv \mu_{\nu_i}/T_{\nu_i} = 3$, such that $T_{\nu_i} \approx \langle E_{\nu_i} \rangle / 3.99$. 
All these values are taken from \cite{Duan:2006an} and are the same as adopted in Refs.~\cite{Lin:2022dbl,Lin:2023nsm,Lin:2024vzy}. 

The BDM flux from an individual SN arriving at Earth at time $t^\prime$ after the SN explosion occurs is computed by integrating over the solid angle $(\theta, \varphi)$ in the Earth-centered spherical coordinate system,
\begin{multline}\label{eq:BDM_flux}
\frac{d\Phi_{\chi}(T_\chi, t^\prime)}{dT_{\chi}} =  \\
   \left.\tau_s\int_0^{2\pi} d\varphi\int_{0}^{\pi/2} d\theta\, \sin\theta \frac{d\Phi_\chi}{dT_\chi d\Omega}(r,d,T_{\chi},\psi)\right|_{t^{\prime}=\frac{d}{c}+\frac{\ell}{v_{\chi}}}. 
\end{multline}
Note that the propagation time $t^\prime$ can be related to $d$ and $\ell$ by
\begin{equation}\label{eq:tp_constraint}
    t^\prime = \frac{d}{c} + \frac{\ell}{v_\chi} 
\end{equation}
with $d=\sqrt{R_s^2+\ell^2 - 2R_s \ell\cos\theta}$.
The integrand in Eq.~\eqref{eq:BDM_flux}, 
\begin{equation}\label{eq:BDM_spatial_dist}
    \frac{d\Phi_\chi}{dT_\chi d\Omega} = \mathcal{J} j_{\chi}(r,d,T_{\chi},\psi)
\end{equation}
is the differential flux on the sky and
\begin{equation}\label{eq:J}
    \mathcal{J}=\left(\frac{\ell - R_s \cos\theta}{cd} + \frac{1}{v_{\chi}}\right)^{-1} 
\end{equation}
is the Jacobian factor.
We approximate $\tau_s$ by 10~s, which is the characteristic SN$\nu$ emission timescale. 
By defining a shifted time variable
\begin{equation}\label{eq:g_t_relation}
    t \equiv t^\prime-t_\nu= \frac{d}{c} + \frac{\ell}{v_\chi} - t_\nu,
\end{equation}
with $t_\nu = R_s / c$ being the SN$\nu$ travel time from $\mathsf{S}$ to $\mathsf{E}$, we align $t = 0$ with the arrival time of SN$\nu$ at Earth.

\begin{figure}
\begin{centering}
\includegraphics[width=0.99\columnwidth]{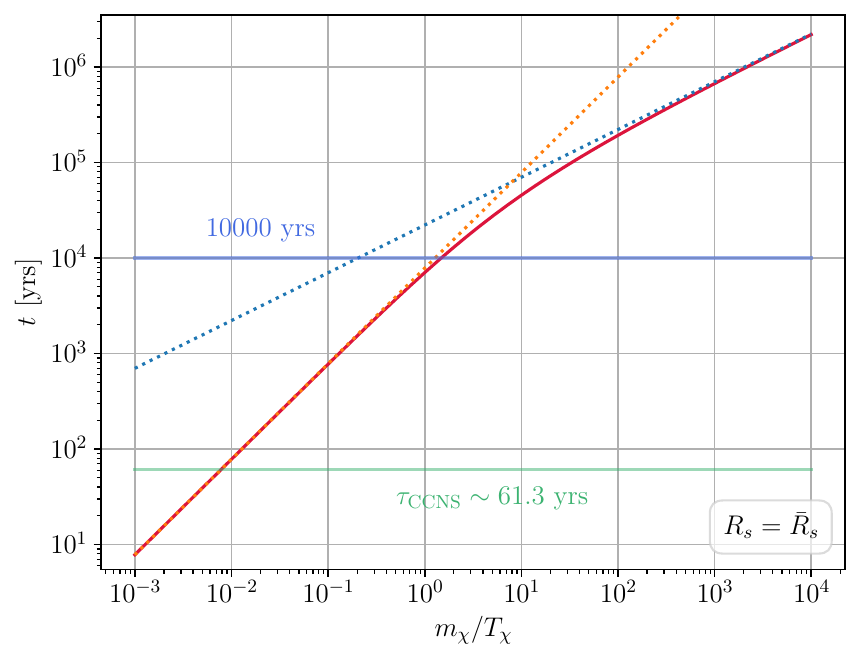}
\end{centering}
\caption{\label{fig:t_van}
$t_{\rm van}$ versus $m_\chi$/$T_\chi$ evaluated at $R_s=\bar{R}_s \approx 9.6$ kpc (red).
Also shown are the approximated solutions of $t_{\rm van}$ from Eq.~\eqref{eq:t_van_approx} in the nonrelativistic (blue dotted line) and relativistic limits (orange dotted line).
Two horizontal solid lines corresponding to 10,000 yrs (blue) and $\tau_{\rm CCNS}$ (green) are shown for reference. }
\end{figure}

As pointed out in Refs.~\cite{Lin:2022dbl,Lin:2023nsm}, the BDM flux from an individual SN lasts for a certain duration $t_{\rm van}$ before vanishing, which can be determined by
\begin{equation}\label{eq:tvan}
    t_{\rm van} (R_s,v_\chi) = \left[\frac{d(\theta)}{c}+\frac{\ell(\theta)}{v_\chi}-t_\nu\right]_{\theta=\theta^*},
\end{equation}
where $\theta^*$ satisfies
\begin{equation}\label{eq:theta*}
\frac{\cos(\psi_{\rm max} - \theta^*)}{\cos\theta^*} = \frac{v_\chi}{c}.
\end{equation}
In Eq.~\eqref{eq:theta*}, 
\begin{equation}\label{eq:psi_max}
    \psi_{\rm max} = \cos^{-1}\left(\frac{T_\chi}{|\mathbf{p}_\chi|}\right) \leq \frac{\pi}{2}
\end{equation}
is the maximum allowed scattering angle and  $|\mathbf{p_\chi}|=\sqrt{T_\chi(T_\chi+2m_\chi )}$. 
In the ultrarelativistic ($m_\chi/T_\chi\ll 1$) and nonrelativistic ($m_\chi/T_\chi \gg 1$) limits, $t_{\rm van}$ can be approximated by (see Appendix~\ref{app:t_van_approx} for derivation) 
\begin{equation}\label{eq:t_van_approx}
t_{\rm van}(R_s,v_\chi)\approx\begin{cases}
\frac{m_\chi}{4T_\chi}\frac{R_s}{c}, &m_\chi/T_\chi \ll 1, \\
\sqrt{\frac{m_\chi}{2T_\chi}}\frac{R_s}{c}, &m_\chi/T_\chi  \gg 1. 
\end{cases}
\end{equation}
In addition to $t_{\rm van}$, another important timescale defined in Refs.~\cite{Lin:2022dbl,Lin:2023nsm} is 
\begin{equation}\label{eq:tp}
     t_p (R_s,v_\chi) =  \frac{R_s}{v_\chi} - t_\nu,
\end{equation}
which corresponds to the delayed arrival time 
of BDM from locations very close to the SN position after the arrival of the SN$\nu$. 

To understand the characteristic $t_{\rm van}$ of SN$\nu$~BDM from the MW for different $v_\chi$ (which can be related to $m_\chi/T_\chi$), we compute numerically 
$t_{\rm van}(\bar R_s,v_\chi)$, shown by the red solid line in Fig.~\ref{fig:t_van}, for a SN that occurs at a distance away from the Earth given by the averaged distance weighted over the spatial distribution function $p_{\rm SN}$ 
\begin{equation}
    \bar{R}_s = \int dR\, dz\, d\phi\, R_s\, p_{\rm SN}(R,z) \approx 9.6~{\rm kpc},
\end{equation}
where $R_s$ is evaluated as
\begin{equation}
R_s = \sqrt{(R_\star\cos\phi_\star+R_e)^2 + R_\star^2 \sin^2\phi_\star + z_\star^2}
\end{equation}
with $(R_\star, z_\star, \phi_\star)$ being the SN location in the galactic coordinate (see Fig.~\ref{fig:cylind_coord}). 
Also shown in Fig.~\ref{fig:t_van} by the dotted lines are the approximate $t_{\rm van}$ evaluated with Eq.~\eqref{eq:t_van_approx}. 
Fig.~\ref{fig:t_van} shows that for the nonrelativistic case with $m_\chi/T_\chi=10$, $t_{\rm van}(\bar R_s)\simeq 4.5\times 10^4$~yrs, implying that many SNe that happened during this time span with $t_{\rm van}\geq t_{\rm age}$ should still contribute to the present-day SN$\nu$~BDM flux. 
However,
for the relativistic case with e.g., $m_\chi/T_\chi=0.01$, $t_{\rm van}(\bar R_s)\simeq 1.1\tau_{\rm CCSN}$, i.e., the duration of the SN$\nu$~BDM is comparable to the averaged interval between successive SNe in MW. 
In this case, the SN$\nu$~BDM at Earth likely originates from very few SNe at any given time.

In Fig.~\ref{fig:SNe_with_non_vanishing_flux}, we indicate the SNe in our sample whose $t_{\rm van} \geq t_{\rm age}$ by filled orange dots shown in the projected $x$-$y$ plane in the galactic coordinate (same as Fig.~\ref{fig:sne_scattering}) for $(T_\chi,m_\chi)=(0.5,5)$ and $(500,5)$ keV, to support the above reasoning. 
The blue dots represent those SNe that do not contribute to the present-day BDM flux. 
For $m_\chi/T_\chi=10$ ($v_\chi\sim 0.4c$), the fraction of SNe with $t_{\rm van}\geq t_{\rm age}$, denoted by $\eta$, is 44.81\% (left figure).
This corresponds to a nonvanishing contribution of BDM from 747 SNe, confirming that a large fraction of past SNe in the MW should contribute to the present-day BDM flux when considering the nonrelativistic $T_\chi$ range. 
In contrast, for relativistic case with $m_\chi/T_\chi=0.01$ (right figure), only 1 SN still produces nonzero BDM flux at Earth ($\eta\simeq 0.06\%$), confirming that when $t_{\rm van}(\bar R_s)\simeq \tau_{\rm CCSN}$, the SN$\nu$~BDM flux is dominated by at most a few SNe.

Since $t_{\rm van}$ depends only on $m_\chi/T_\chi$, in what follows, we fix $m_\chi=5$ keV and take 
different $T_\chi$ values between 0.5~keV and 5~MeV to explore the
dependence of the
temporal and angular profiles of SN$\nu$~BDM from our SN sample 
on $m_\chi/T_\chi$.

\section{Numerical results}\label{sec:numerical_results}

\begin{figure*}
\begin{centering}
\includegraphics[width=0.9\columnwidth]{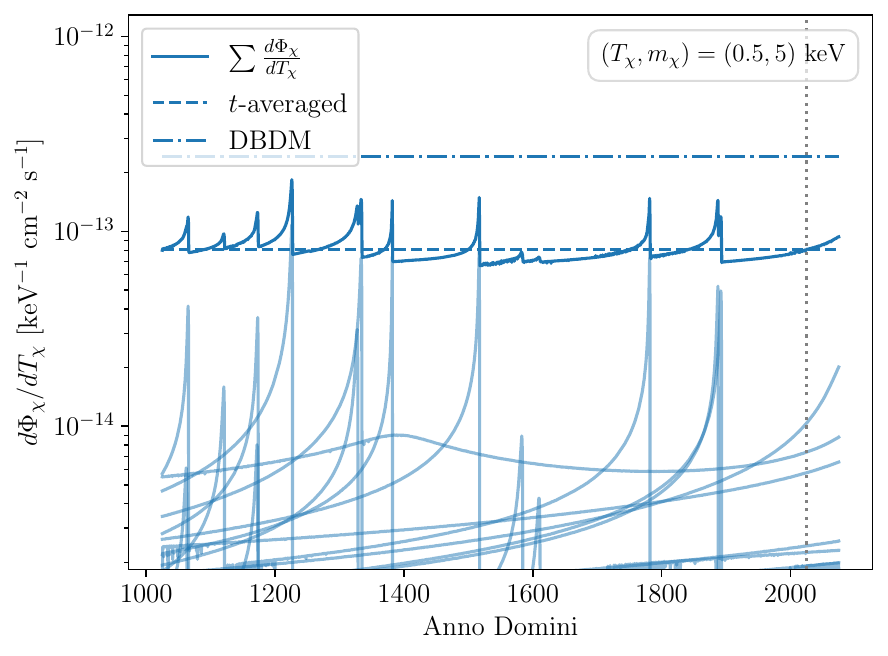}
\includegraphics[width=0.9\columnwidth]{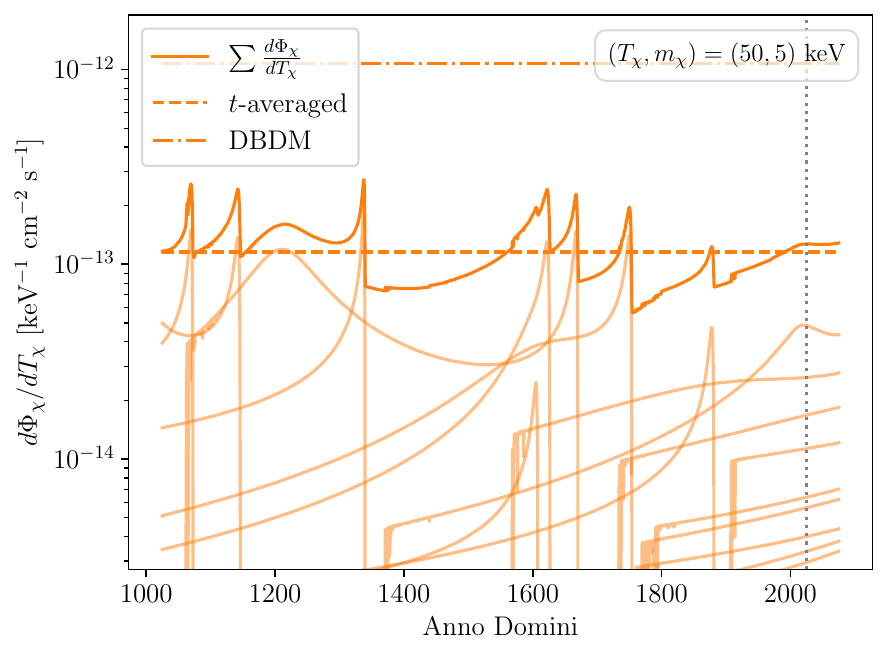}\\
\includegraphics[width=0.9\columnwidth]{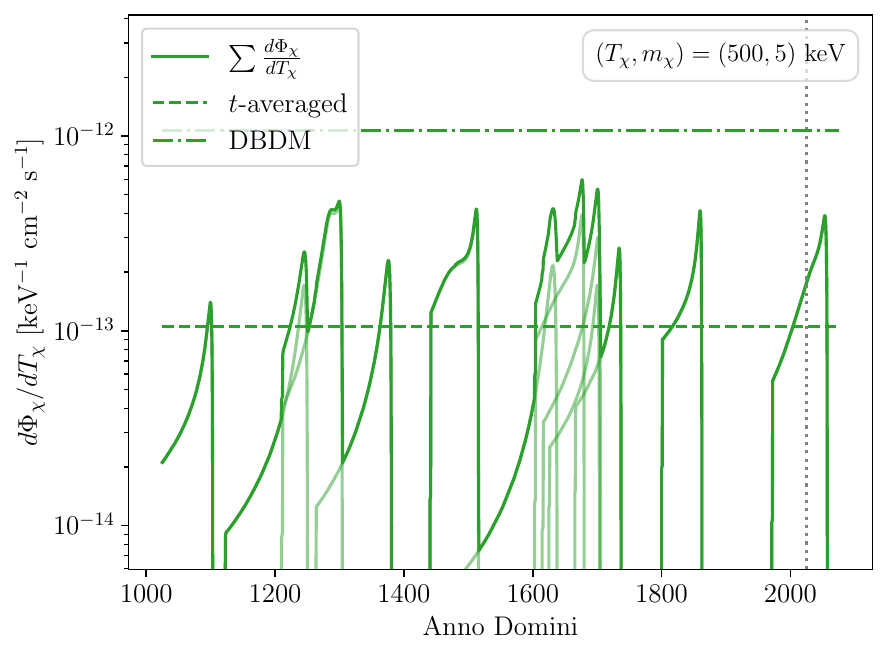}
\includegraphics[width=0.9\columnwidth]{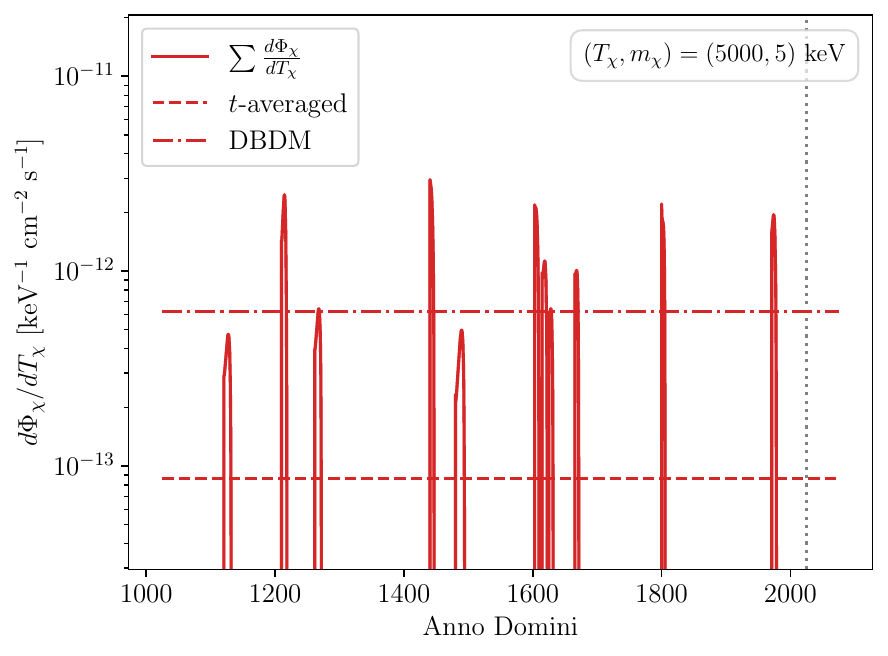}
\end{centering}
\caption{\label{fig:BDM_flux_millennium}
Evolution of BDM fluxes from MW SNe in our mocked sample over a time span of one millennium. 
We take $m_\chi = 5$ keV, with different values of $T_\chi$ indicated in the top right corner of each panel.  
The solid lines show the MW cumulative BDM flux (labeled by $\sum \frac{d\Phi_\chi}{dT_\chi}$) obtained by summing contributions from all individual MW SNe (light solid lines). 
The dashed lines represent the time-averaged BDM flux over the same duration shown in the plot.
The dot-dashed lines show the DBDM flux from all SNe in galaxies at higher redshifts.
Gray vertical dotted lines mark the present day, assuming year 2025.
}
\end{figure*}

In the previous sections, we discussed how SNe are distributed in the MW according to $p_{\rm SN}$, and what fraction of them can contribute to a nonzero BDM flux at the present day.
In this section, we first evaluate the temporal evolution of the BDM flux over the last $\sim\mathcal{O}(10^3)$~yrs by summing over all SN contributions, each of which being evaluated with Eq.~\eqref{eq:BDM_flux} in Sec.~\ref{sec:flux_temp}.  
In Sec.~\ref{sec:spatial_dist}, we further examine how present-day BDM flux is distributed across the celestial sphere using Eq.~\eqref{eq:BDM_spatial_dist}.
In evaluating the fluxes from each SN, we have placed a minimal cutoff for $d$ at $d_{\rm cut}=200$~km. 
While the choice of $d_{\rm cut}$ here is much smaller than the cutoffs chosen in Refs.~\cite{Lin:2022dbl,Lin:2023nsm,Lin:2024vzy}, we have verified that a different choice only results in negligible changes to the results presented here.

\subsection{Evolution of BDM flux}\label{sec:flux_temp}

In Fig.~\ref{fig:BDM_flux_millennium}, we show the BDM flux from each MW SN (curves with lighter colors) computed with Eq.~\eqref{eq:BDM_flux} and the total flux summing over all MW SNe in our sample (darker colors; labeled by $\sum\frac{d\Phi_\chi}{dT_\chi}$), spanning from \textsc{ad}~1025 to \textsc{ad}~2075 for four representative values of $T_\chi=0.5, 50, 500$, and 5000~keV.  
The present day, assumed as year 2025, is marked by the gray dotted line. 
For each $T_\chi$, we also plot the DBDM flux computed in Ref.~\cite{Lin:2024vzy} by the dot-dashed line. 
From here onward, unless explicitly stated it is DBDM, the acronym BDM refers to those coming from the SNe in MW. 

For $T_\chi = 0.5$ keV shown in the top left panel, corresponding to the same case shown in the left panel of Fig.~\ref{fig:SNe_with_non_vanishing_flux}, there are more than 700 SNe contributed to the present-day BDM flux since 
the average vanishing time of all 1,667 SNe in our dataset
$\bar t_{\rm van}\simeq 5.3 \times 10^4$ years.\footnote{This value is similar to the characteristic value of $t_{\rm van}(\bar R_s)$ evaluated for a representative SN at $\bar R_s$ reported in the previous section for the same $m_\chi/T_\chi$.} 
However, even in this case, the total BDM flux still varies in time by a factor of $\sim 3$.
This is mainly because the BDM flux from each individual SN peaks close to $t_{\rm van}$, as clearly shown by curves with lighter colors.
It is important to note that at low $T_\chi$ range with $T_\chi\ll\langle E_{\nu_i}\rangle$, the temporal behavior of the BDM flux is different from that for $T_\chi\gtrsim\langle E_{\nu_i}\rangle$, where the flux peaks at $\sim t_p$ when $\sigma_{\chi\nu}$ is energy-independent ~\cite{Lin:2022dbl,Lin:2023nsm}. 
We find that this is because when $T_\chi\ll\langle E_{\nu_i}\rangle$, BDM that arrives at later times are dominantly upscattered at locations closer to where the SN explodes. 
Meanwhile, the required neutrino energy for BDM stays nearly constant over time. 
Hence, $dn_\nu/dE_\nu\propto d^{-2}$ increases with time, leading to a larger BDM flux closer to $t_{\rm van}$.
On the other hand, when $T_\chi\gtrsim \langle E_{\nu_i} \rangle$, the required neutrino energy for BDM arriving at $t\gtrsim t_p$ increases with time and can largely exceed $\langle E_{\nu_i} \rangle$. 
In this case, the exponential suppression of $dn_\nu/dE_\nu$ at $E_\nu\gg \langle E_{\nu_i} \rangle$ results in a decrease of the BDM flux for $t\gtrsim t_p$ reported in Refs.~\cite{Lin:2022dbl,Lin:2023nsm}.
More importantly, the BDM flux is not dominated by any single supernova, aside from the peaks. 
The BDM flux outside the short durations around the peaks differs from the time-averaged flux only by $\lesssim 15\%$ and may be approximated as a stationary component.  
In particular, the largest contribution from a single SN to the present-day BDM flux is only up to $\sim 13\%$.

As $T_\chi$ increases, the number of SNe contributing to the cumulative BDM flux decreases rapidly. 
The overlaps of BDM fluxes from individual SNe diminish, resulting in a less steady total flux. 
For example, with $m_\chi/T_\chi=0.1$ shown in the top right panel of Fig.~\ref{fig:BDM_flux_millennium}, which corresponds to $\bar t_{\rm van}=7.8\times 10^2$~yrs,  
our SN sample yields BDM flux that varies in time by a factor of $\sim 5$. 
At any given time, the total flux is dominated by only a few SNe such that the actual flux can differ from the time-averaged flux by a factor of $\sim 2$.
With an even smaller value of $m_\chi/T_\chi=0.01$ (bottom left panel), which corresponds to $\bar t_{\rm van}\approx 79$~yrs~$ \approx 1.29\tau_{\rm CCSN}$, the BDM flux generally differs from the time-averaged value by up to a factor of $\sim\mathcal{O}(10)$ for most of the times, and can even completely vanish at certain intervals.
Finally, for the ultrarelativistic limit with $\bar t_{\rm van}\approx 7.9~{\rm yrs}\approx 0.13 \tau_{\rm CCSN}$, $t_{\rm van}$ from individual SN only last for weeks to a few years after SN$\nu$ arrives.
Thus, the BDM flux becomes nonzero only around the time when a SN explodes. 

The above discussions suggest that only when 
$t_{\rm van} \gg \tau_{\rm CCSN}$, which corresponds to $m_\chi/T_\chi\gtrsim \mathcal{O}(10)$, may the resulting BDM flux at Earth be approximated as steady.
In the next subsection, we will show that similar criteria also apply to the angular distribution of BDM flux over the sky.
Moreover, our results here show that the SN$\nu$~BDM flux only exceeds the DBDM flux in the ultrarelativistic case when $t_{\rm van}\ll \tau_{\rm CCSN}$, indicating that the diffuse component of SN$\nu$~BDM should be dominated by the contribution from all SNe at higher redshifts, but not from the past SNe in MW.

\begin{figure*}
\begin{centering}
\includegraphics[width=0.99\columnwidth]{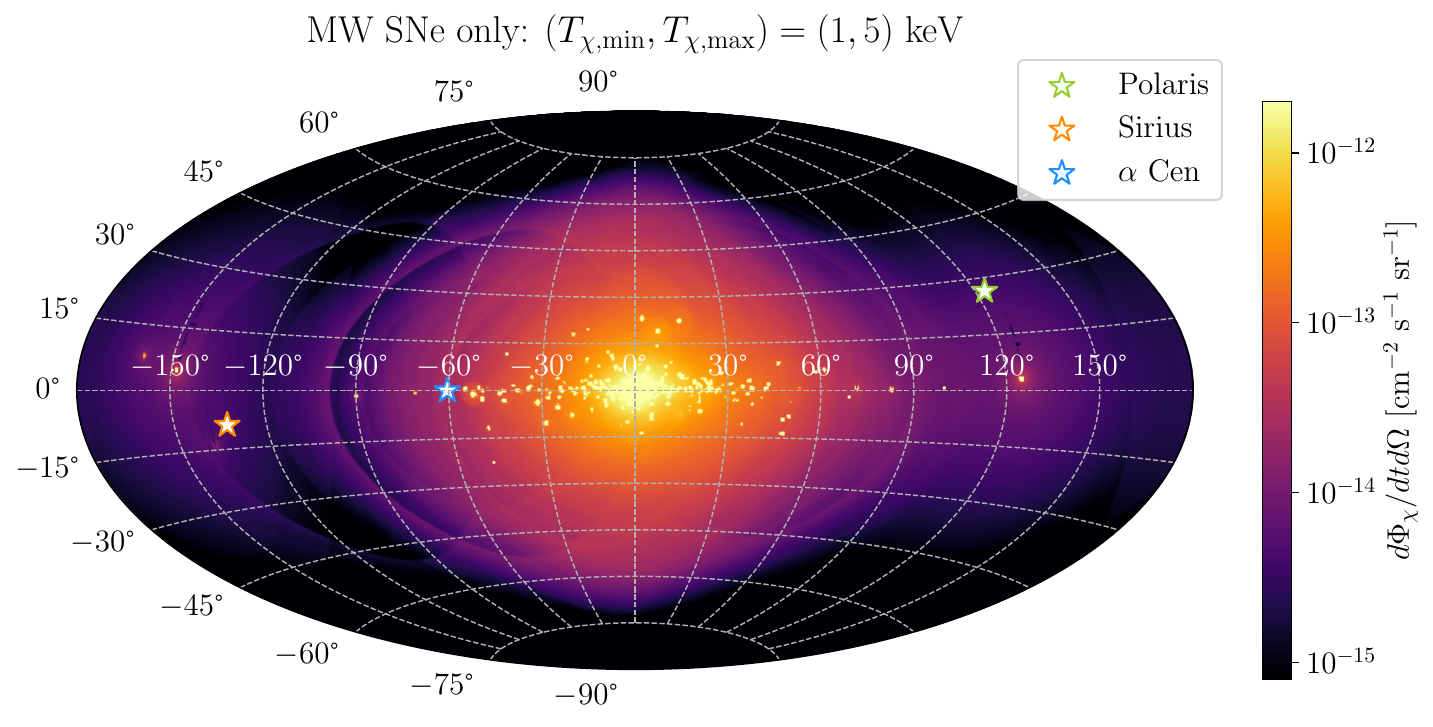}
\includegraphics[width=0.99\columnwidth]{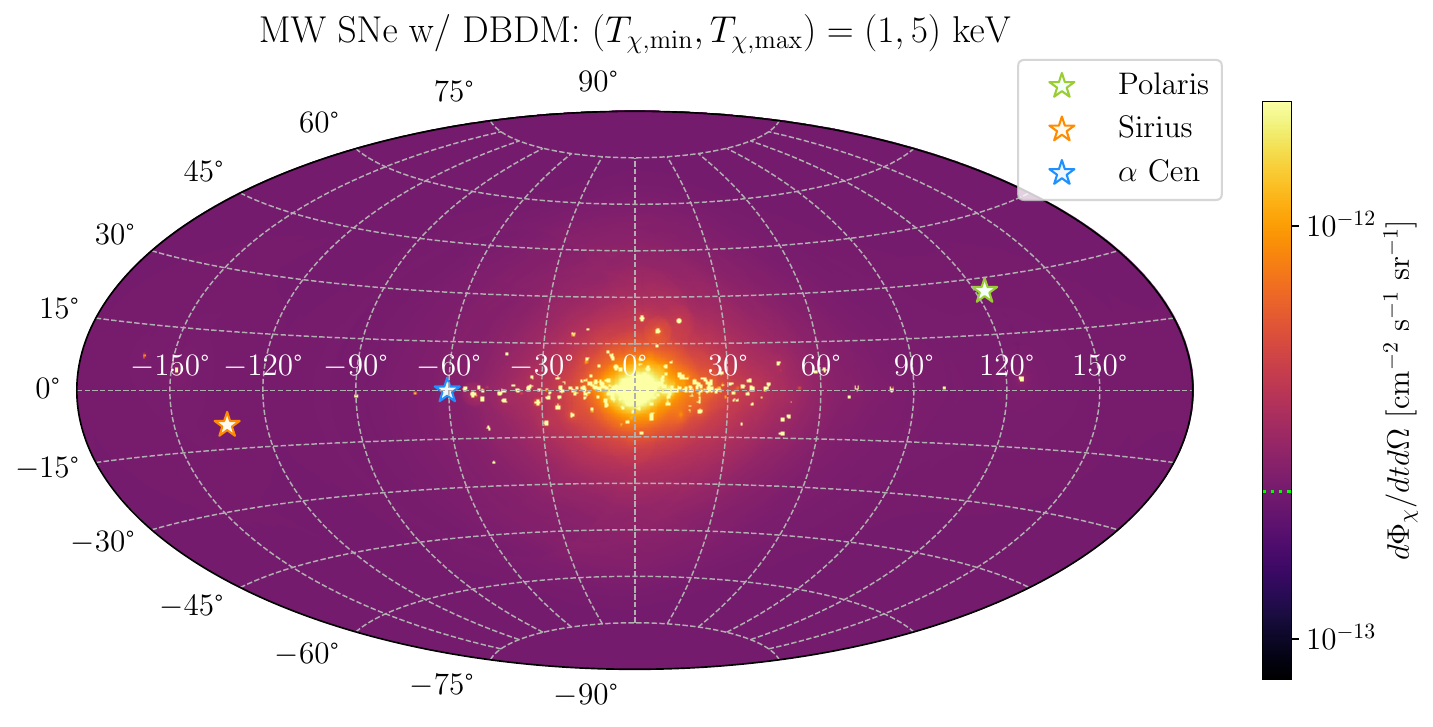}\\
\includegraphics[width=0.99\columnwidth]{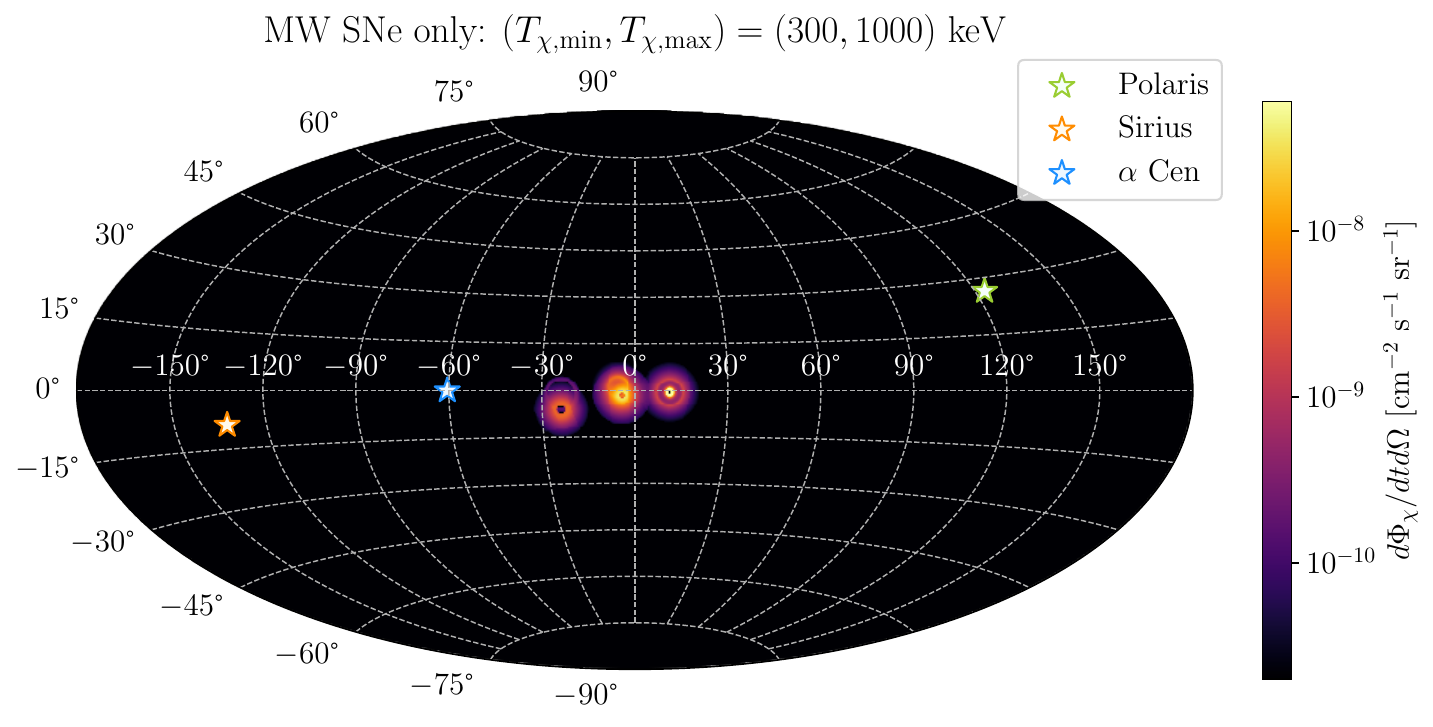}
\includegraphics[width=0.99\columnwidth]{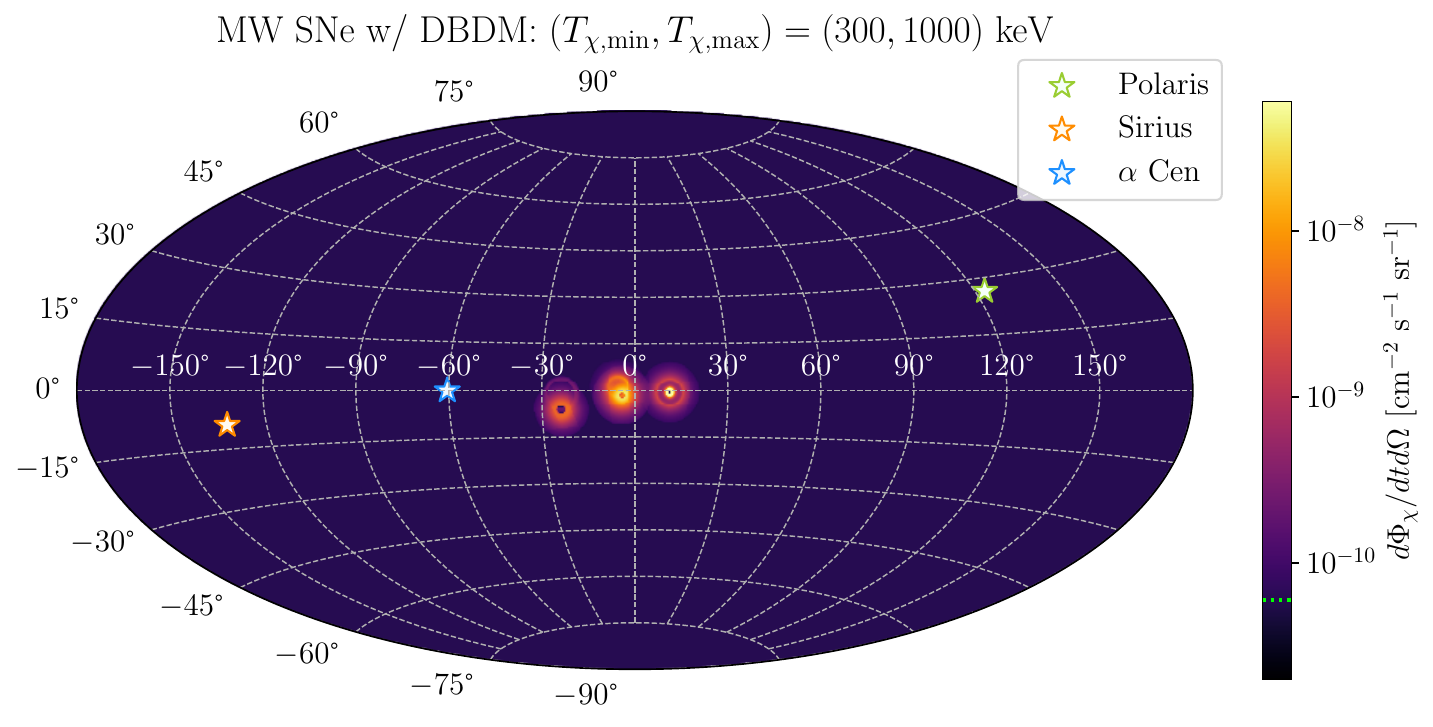}
\end{centering}
\caption{\label{fig:allsky_map}
$T_\chi$-integrated BDM flux per steradian over the celestial sphere, in galactic coordinate, at present day for different $T_\chi$ ranges indicated by the labels.
Left panels show the result that only considers MW SNe from our mocked dataset, while the right panels include contributions of DBDM, whose values are indicated by the short green dotted lines in the color bar ($2.28\times 10 ^{-13}$ and $5.98\times 10 ^{-11}$ cm$^{-2}$ s$^{-1}$ sr$^{-1}$ in the upper and lower panels, respectively).
We assume $m_\chi=5$ keV for both cases. The locations of three representative stars, Polaris, Sirius and $\alpha$~Centauri are marked by star symbols.} 
\end{figure*}

\begin{figure}
\begin{centering}
\includegraphics[width=0.9\columnwidth]{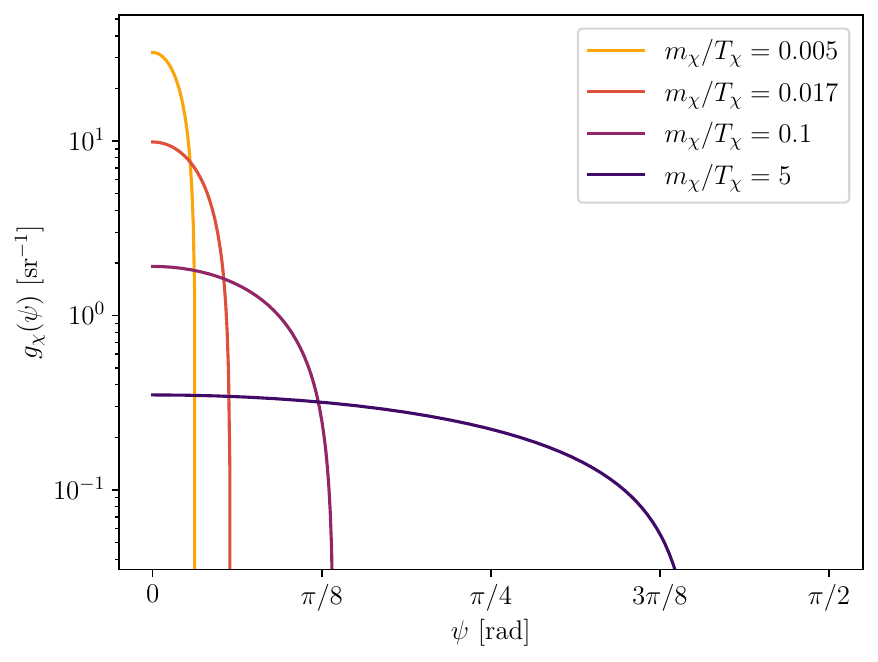}
\end{centering}
\caption{\label{fig:ang_dist}
The BDM angular distribution at the boosted point ($\mathsf{B}$) for different $m_\chi/T_\chi$ values.
}
\end{figure}

\subsection{BDM angular distribution across the celestial sphere}\label{sec:spatial_dist}

After examining the temporal profile of angle-integrated SN$\nu$~BDM flux, we now investigate their angular distribution over the sky. 
We evaluate Eq.~\eqref{eq:BDM_spatial_dist} and integrate over different $T_\chi$ ranges by 
\begin{equation}\label{eq:Tx_integrated_BDM_flux}
\frac{d\Phi_{\chi}(t)}{d\Omega} =
\left. \tau_s \int_{T_{\chi,{\rm min}}}^{T_{\chi,{\rm max}}} dT_\chi\,
\frac{d\Phi_\chi}{dT_\chi d\Omega}
\right|_{t = \frac{d}{c} + \frac{\ell}{v_{\chi}} - t_\nu}
\end{equation}
at any given time $t$.
We fix $m_\chi=5$~keV and consider two different ranges for $T_\chi$ for demonstration: $(T_{\chi,{\rm min}}, T_{\chi,{\rm max}})=(1,5)$ 
and $(300,1000)$~keV. 
Left panels of Fig.~\ref{fig:allsky_map} show the resulting distributions at present day ($t=10^5$ years) considering the MW SN$\nu$~BDM only.  
For reference, we also mark the angular position of three stars: Polaris, Sirius, and $\alpha$~Centauri, in each panel. 
Clearly, with low $T_\chi$ range in the nonrelativistic limit (upper left panel), the resulting BDM fluxes from individual SNe become very dispersive and extend to higher latitudes.
This can be understood by examining the BDM angular distribution $g_\chi(\psi)$ versus the scattering angle $\psi$ at a given boosted point ($\mathsf{B}$) [Eq.~\eqref{eq:gx}] for different values of $m_\chi/T_\chi$ shown in Fig.~\ref{fig:ang_dist}. 
Clearly, the larger the $m_\chi/T_\chi$ is, the bigger the spread of the BDM flux across the sky.
As a result, contributions from different SNe largely overlap, which gives rise to an overall larger flux toward the direction of the GC.  
Although the figure also shows that the BDM flux peaks within a very small solid angle range centered at directions corresponding the location of contributing SNe, 
due to the larger value of $dn_\nu/dE_\nu$ close to the SN location [see Eq.~\eqref{eq:dnv/dEv}], 
it would require an extremely good angular resolution in experiments to isolate them. 
Thus, the BDM flux in this limit can be practically considered as spatially diffusive.
On the other hand, for $300~{\rm keV}\leq T_{\chi}< 1000$~keV (bottom left panel), only 5 SNe close to the GC contribute to the BDM flux.
If an angular resolution of a few degrees can be achieved, it is possible to separate contributions from individual SNe.

Note that the angular distribution of BDM from each SN exhibits a ringlike feature in both cases. 
This is 
related to the fact that BDM with larger scattering angle $\psi$ is upscattered by neutrinos with 
higher $E_\nu$; see Eq.~(A10) in Ref.~\cite{Lin:2023nsm}.  
For $T_\chi<1$~MeV$\ll \langle E_{\nu_i} \rangle$ considered in Fig.~\ref{fig:allsky_map},  
$dn_\nu/dE_\nu$ increases with $E_\nu^2$, resulting in larger BDM flux at larger $\psi$. 
The ringlike shape is clearly visible in the bottom left panel where the contributions of each SN are separated, but becomes obscured in the upper left panel at the diffuse limit. 

In the right panels of Fig.~\ref{fig:allsky_map}, we show the expected angular distributions that include the DBDM component. 
Clearly, for the case with nonrelativistic $T_\chi$ range shown in the upper right panel, the contribution from the MW SNe only stands out from the isotropic DBDM flux within $\sim 10^\circ$ in latitude and $\sim 30^\circ$ in longitude around the GC. 
Outside this angular range, the DBDM flux dominates. 
For the relativistic case (lower right panel), the BDM fluxes from individual SNe are generally larger than the DBDM flux. 
However, as the all-sky-integrated flux is dominated by DBDM (Fig.~\ref{fig:BDM_flux_millennium}), it would also require good detection angular resolution at these $T_\chi$ ranges to uncover the BDM contribution due to the past MW SNe.

\section{BDM from Supernova Remnants}\label{sec:histroical_SNRs}

\begin{figure*}
\begin{centering}
\includegraphics[width=0.9\columnwidth]{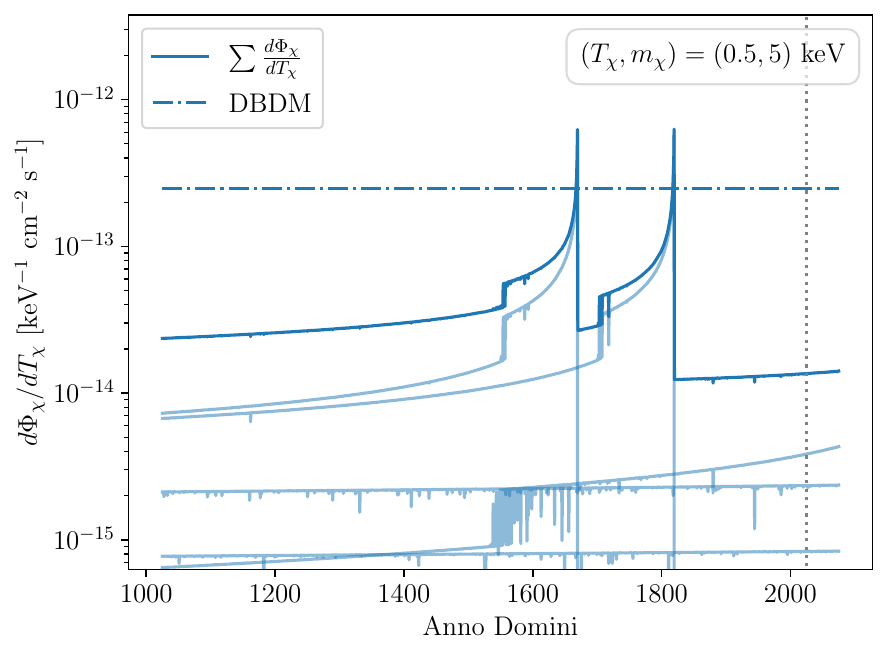}
\includegraphics[width=0.9\columnwidth]{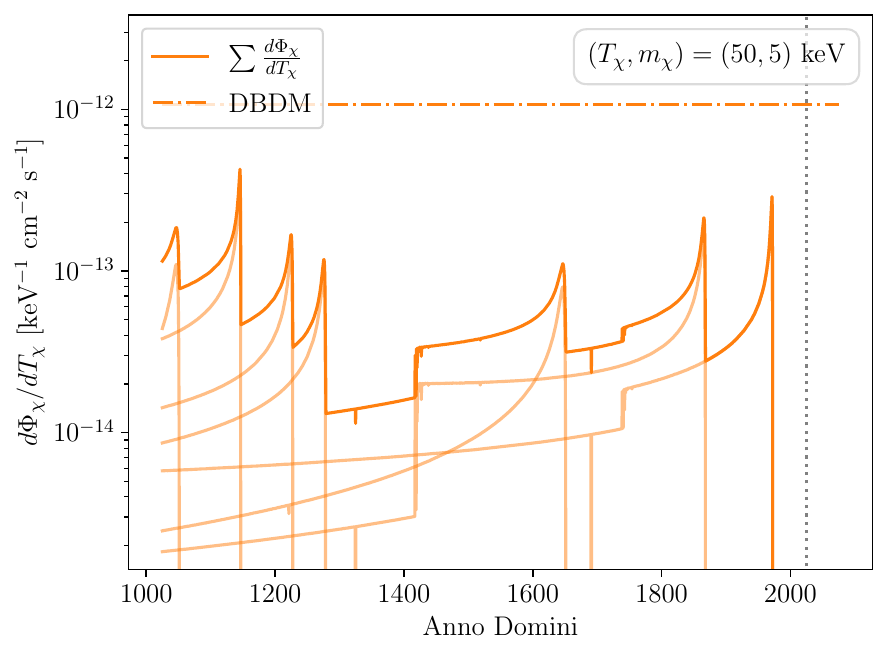}
\end{centering}
\caption{\label{fig:documentedSNR_temporal}
Time evolution of BDM fluxes from 
SNe possibly associated with known SNRs
over a time span of one millennium (see text for details).
The same line styles of Fig.~\ref{fig:BDM_flux_millennium} are used to denote quantities shown in the plots.
}
\end{figure*}

\begin{figure*}
\begin{centering}
\includegraphics[width=0.99\columnwidth]{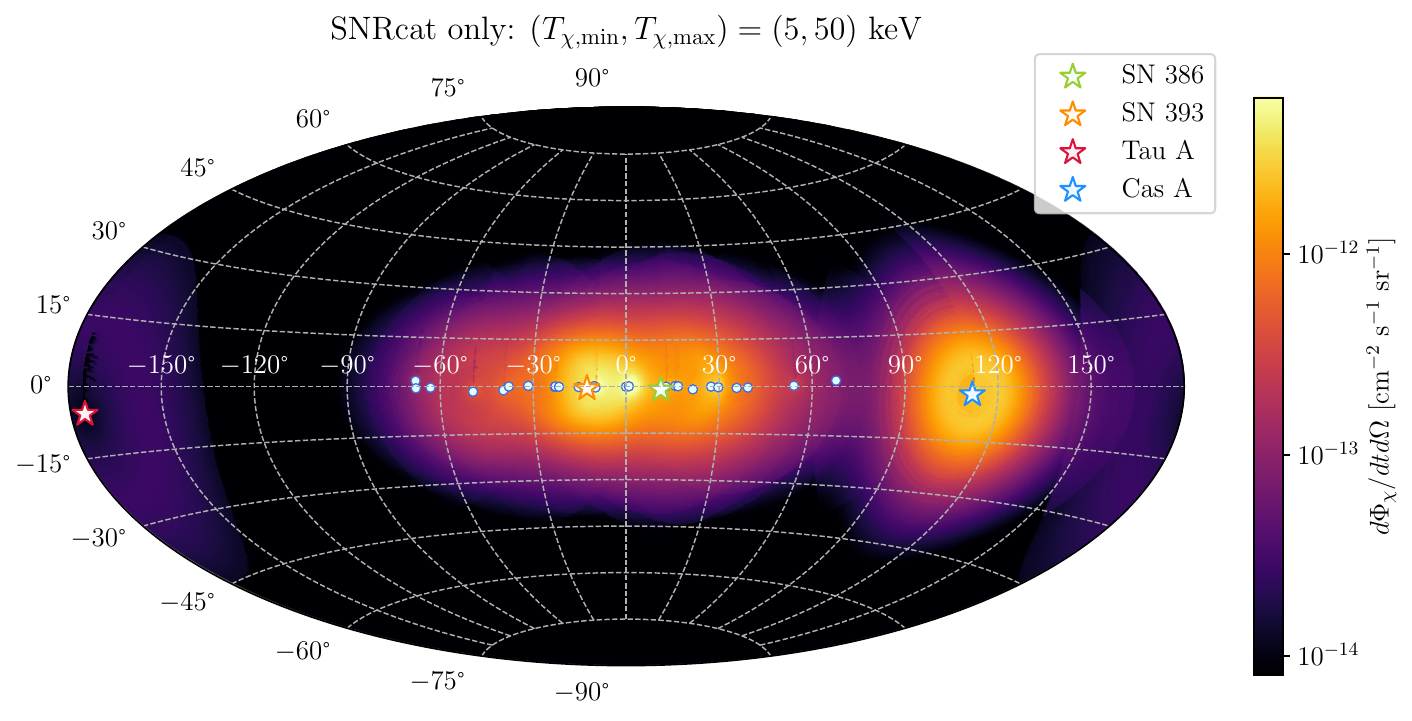}
\includegraphics[width=0.99\columnwidth]{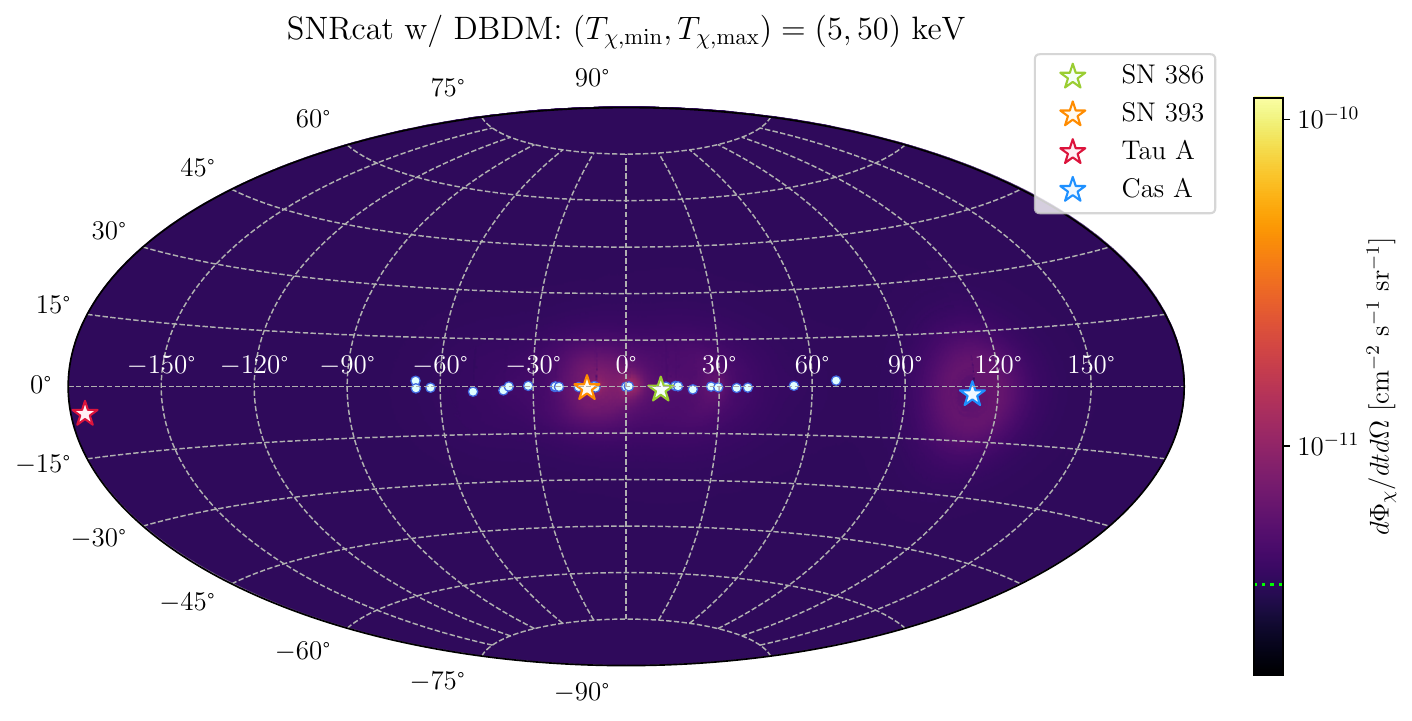}
\end{centering}
\caption{\label{fig:documentedSNR_allsky}
$T_\chi$-integrated BDM flux per steradian over the celestial sphere, in galactic coordinate, at present day, from SNe possibly associated with known SNRs for $m_\chi=5$~keV. 
The left and the right panels show the result that only considers MW SNe and the values including DBDM contribution (indicated by the green dotted line with a value of $3.76 \times 10^{-12}$ cm$^{-2}$ s$^{-1}$ sr$^{-1}$ on the color bar of the right panel), respectively.
Locations of SNe that contribute to the BDM flux are labeled by the blue dots.  
Four known SNe in recent history are marked by star symbols.}  
\end{figure*}

So far we have focused on results obtained using the mocked SN dataset shown in 
Fig.~\ref{fig:sne_scattering}.
However, hundreds of SNRs have been observed and documented, see e.g., SNRcat~\cite{SNRcat} for a catalog.  
Here, we analyze the temporal and spatial BDM distributions resulting from this catalog with the provided SNR age, distance, and direction estimations as well as their associated remnant properties. 
Because not all listed SNRs contain all the above information needed to evaluate the corresponding BDM flux, we only consider SNRs with: 
(1) potential association with remnants classified as neutron star (NS), pulsar remnant (PSR), or pulsar wind nebula (PWN), and  
(2) identifiable age and distance.  
Given the large uncertainties in inferring the ages and distances of SNRs, we take the average of the reported minimum and maximum values as their true ages and distances.  
If only one single value is provided for age or distance, we use it as the nominal one.  
After this selection, 81 SNRs remain in this ``realistic'' SN dataset. 
Among them, 77 SNRs have ages less than 100,000~yrs, with distances 
between 0.6 and 16.7 kpc.
We then perform the same calculations to obtain the corresponding temporal and spatial BDM fluxes at the present day, which are shown in Figs.~\ref{fig:documentedSNR_temporal} and \ref{fig:documentedSNR_allsky}, respectively. 
Note that the number of SNe in this dataset is much smaller than that of the mock data mainly due to dust distinction as well as observational limits. 
Thus, the results presented below can be considered as the ``floor'' for BDM from the MW's past SNe. 

For the temporal distribution shown in Fig.~\ref{fig:documentedSNR_temporal}, we again take $T_\chi = 0.5$ (left panel) and 50 keV (right panel) and fix $m_\chi=5$~keV.
Here, even with $m_\chi/T_\chi=10$, the cumulative BDM flux does not exhibit the diffuse nature shown in the upper left panel of Fig.~\ref{fig:BDM_flux_millennium}, due to the limited number of identified SNRs.
Except the two specific peaks between \textsc{ad} 1650 and \textsc{ad} 1850 associated with very nearby SNRs,\footnote{These two distinct peaks 
are from Vela Jr.~(G266.2--01.2) and Boomerang (G106.3+02.7), both of which are located approximately 0.75 kpc from Earth and are estimated to be 3750 and 3900 year-old, respectively.} the flux is much smaller than the DBDM one.
For $T_\chi=50$~keV (right panel), the BDM flux sits completely under the DBDM flux all the time and is zero now. 

For the BDM spatial distribution (Fig.~\ref{fig:documentedSNR_allsky}), integrated over a $T_\chi$ range within $(T_{\chi,{\rm min}}, T_{\chi,{\rm max}}) = (5, 50)$ keV, the DBDM contribution is excluded (included) in the left (right) panel.
The locations of SNe whose contributions are nonzero are marked by blue dots. 
Four specific remnants: SN~386 (G011.2--00.3), SN~393 (G347.3--00.5), Taurus A (G184.6$-$05.8, also know as the Crab Nebula), and Cassiopeia A (G111.7--02.1) are particularly labeled by star symbols for reference. 
With this $T_\chi$ range, the BDM fluxes from each SN spread widely over the sky and overlap with each other. 
However, once the DBDM flux is considered, the BDM fluxes are mostly buried under the DBDM, except for those around Taurus A and Cassiopeia A.
We do not show results in a larger $T_\chi$ range. 
This is because in that case, the ages of all SNe in this dataset
are greater than their $t_{\rm van}$ such that they cannot contribute to the present-day flux.

\section{Summary}\label{sec:summary}

In this work, we have investigated the temporal and spatial distribution of the SN$\nu$~BDM flux for BDM with low kinetic energy $T_\chi$ such that the typical BDM flux duration, $t_{\rm van}$, from a single galactic SN is comparable to or larger than the typical time interval between two successive SNe.
Due to insufficient information of all SNe in the MW history, we rely on MC simulation to generate a mock dataset that contains the locations and ages of 1,667 SNe over a time span of 100,000~yrs. 
Using this dataset, we computed the detailed time evolution of the angle-integrated BDM flux for different values of $m_\chi/T_\chi$ covering both the nonrelativistic and the relativistic limits. 
We find that for nonrelativistic BDM whose $m_\chi/T_\chi=10$, BDM from over $44\%$ of SNe in our dataset still contribute to the present-day flux. 
We have also shown that in this case, the accumulated SN$\nu$~BDM varies within $\sim 15\%$ of their time-averaged value, except for short durations where the flux from individual SN peaks. 
Thus, most of the time, the time-averaged SN$\nu$ BDM flux can be taken as a good approximation. 
However, for relativistic SN$\nu$~BDM with $m_\chi/T_\chi<0.1$, corresponding to a typical $t_{\rm van}>\mathcal{O}(800)$~yrs, 
the variation of their temporal flux is generally large and cannot be well approximated by steady-state approximation. 
This implies that when considering the SN$\nu$~BDM with sub-MeV $T_\chi$ relevant to, e.g., semiconductor-based DM searchers, as done in Ref.~\cite{Sun:2025gyj}, this effect can become important and cannot be neglected in determining the BDM flux originated from the past SNe in MW. 

We have also evaluated the angular distribution of the same SN$\nu$~BDM flux at present.
We find that it can be considered as diffuse for the nonrelativistic BDM, since the contribution of individual SNe are dispersed over a large sky area and overlap with each other. 
On the other hand, the BDM flux from different SNe may be discerned given enough angular resolution in the relativistic regime.  
For completeness, we have also utilized available SNR data, which have about 80 SNRs potentially associated with CCSNe to estimate the corresponding BDM flux. 

Comparing the BDM flux from past SNe in the MW to the DBDM flux originated from all SNe in galaxies at higher redshifts investigated in Ref.~\cite{Lin:2024vzy}, we find that for $m_\chi/T_\chi\gtrsim 0.01$, the all-sky-integrated MW SN$\nu$~BDM flux is generally smaller than the DBDM flux by a factor of a few.  
Although the MW BDM flux in certain angular domains can be larger than the isotropic DBDM flux, it would require relevant experiments to achieve an angular solution of $\mathcal{O}(10)$ degrees to uncover the MW SN$\nu$~BDM flux with sub-MeV kinetic energy. 
This finding suggests that DBDM should be considered as the primary target for relevant searches of SN$\nu$~BDM before the next galactic SN takes place. 
When such a once-in-a-lifetime event happens, the transient feature of SN$\nu$~BDM at $T_\chi\gtrsim \mathcal{O}(1)$~MeV originally proposed in Ref.~\cite{Lin:2022dbl} will possibly allow various neutrino and DM experiments to better probe the nature of DM, complementary to many other ways.

\begin{acknowledgments}
We thank Bin Zhu for useful discussions.
Y.H.L.~and M.R.W.~acknowledge support from the National Science and Technology Council, Taiwan under Grant No.~111-2628-M-001-003-MY4, the Academia Sinica under Project No.~AS-IV-114-M04, and Physics Division, National Center for Theoretical Sciences of Taiwan. 
We would like to also acknowledge the use of computational resources provided by the Academia Sinica Grid-Computing Center.
\end{acknowledgments}

\bibliography{main}

\appendix

\section{Impact due to different underlying SN spatial distributions}\label{app:H2_density}

\begin{figure}
\begin{centering}
\includegraphics[width=0.9\columnwidth]{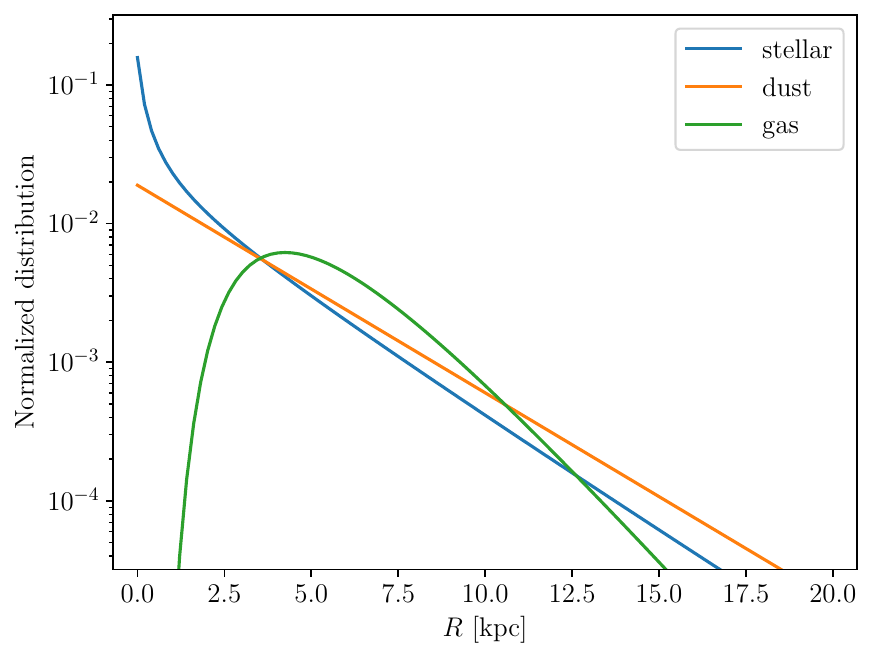}
\end{centering}
\caption{\label{fig:normalized_density_profile}
Normalized height-integrated, radial distributions of three mass components of MW: stellar, dust and H$_2$ gas, based on Eq.~\eqref{eq:p_SN} \cite{Adams:2013ana} and Eq.~\eqref{eq:h2_dist}, respectively.}
\end{figure}

To explore the dependence of results presented on the underlying spatial distribution of CCSNe, we consider two additional cases in this appendix. 
Fig.~\ref{fig:normalized_density_profile} compares the normalized, height-integrated radial distributions of the MW dust \cite{Adams:2013ana} and the molecular hydrogen (H$_2$) gas~\cite{McMillan:2016jtx}, to that of the stellar mass adopted in the main text.
While the dust distribution is similar to that of the stellar mass, the H$_2$ gas distribution is highly suppressed within $\sim 2.5$~kpc. 
Hence, we perform additional calculations for the 
local MW BDM under the diffuse limit with  $(T_\chi,m_\chi)=(0.5,5)$~keV, using the 
H$_2$ gas distribution given by \cite{McMillan:2016jtx}
\begin{equation}\label{eq:h2_dist}
    \rho_g(R,z) = \frac{\Sigma_0}{4z_d}\exp\left(-\frac{R_m}{R}-\frac{R}{R_d}\right){\rm sech}^2\left(\frac{z}{2z_d}\right), 
\end{equation}
where $(R_m,R_d,z_d)=(12,1.5,0.045)$ kpc and $\Sigma_0=2.18\times 10^9~M_\odot\,{\rm kpc}^{-2}$ to replace the stellar distribution profiles used in Eq.~\eqref{eq:p_SN} of the main text.

\begin{figure}
\begin{centering}
\includegraphics[width=0.9\columnwidth]{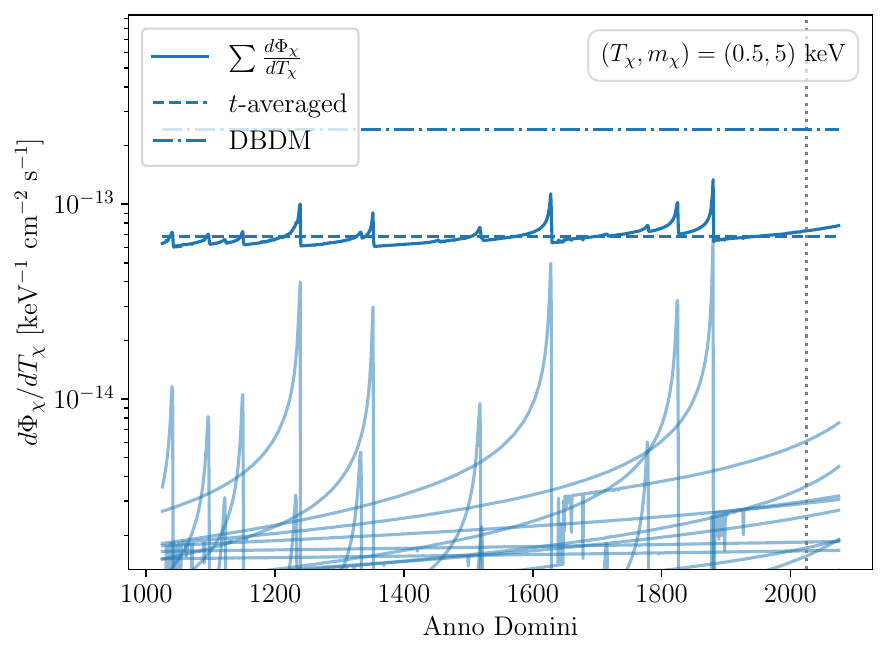}
\end{centering}
\caption{\label{fig:BDM_flux_with_H2}
Evolution of BDM fluxes from MW SNe over a time span of one millennium assuming that CCSN spatial distribution follows the H$_2$ gas given in Eq.~\eqref{eq:h2_dist}.  Legends are identical to Fig.~\ref{fig:BDM_flux_millennium}.}
\end{figure}

With a representative set of CCSNe consisting of 1,593 SNe for this case, the associated temporal evolution of BDM flux over the past millennium is
shown in Fig.~\ref{fig:BDM_flux_with_H2}.  
Compared to the top left panel of Fig.~\ref{fig:BDM_flux_millennium}, the resulting BDM flux has qualitatively similar temporal variation.
The time-averaged flux (dashed) here is  $\simeq 6.85\times 10^{-14}$ ${\rm keV}^{-1}\,{\rm cm}^{-2}\,{\rm s}^{-1}$, which is $\sim 15\%$ smaller than the value derived using the stellar density distribution ($\simeq 8.09\times 10^{-14}~{\rm keV}^{-1}\,{\rm cm}^{-2}\,{\rm s}^{-1}$), illustrating the robustness of our result.  

\section{Impact of DM spike}\label{app:DM_spike}

\begin{figure}
\begin{centering}
\includegraphics[width=0.9\columnwidth]{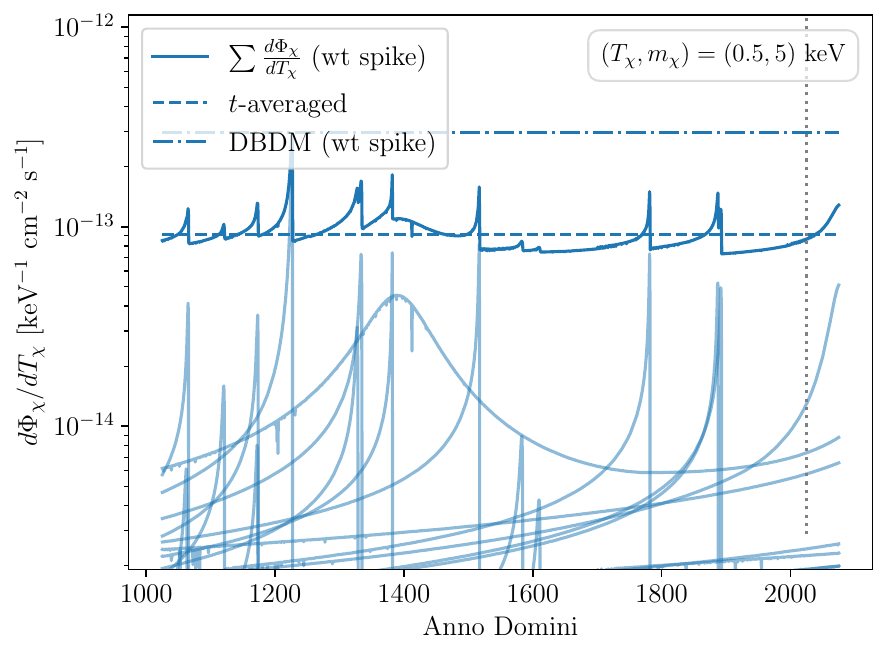}
\end{centering}
\caption{\label{fig:BDM_flux_with_spike}
Evolution of BDM fluxes from MW SNe over a time span of one millennium with DM spike. The SNe sample is the same as shown in Fig.~\ref{fig:sne_scattering}.  Legends are identical to Fig.~\ref{fig:BDM_flux_millennium}.}
\end{figure}

In this appendix, we discuss the impact due to the presence of DM spike \cite{Gondolo:1999ef,Ullio:2001fb,Cline:2022qld} on the BDM flux in the diffuse limit.
We take the same DM spike profiles adopted in Ref.~\cite{Lin:2023nsm} and compute the corresponding MW~SN$\nu$~BDM and the DBDM fluxes. For the MW~BDM, we use the same SN dataset 
shown in Fig.~\ref{fig:SNe_with_non_vanishing_flux}.
The resulting temporal evolution of MW BDM flux as well as the DBDM flux are displayed in Fig.~\ref{fig:BDM_flux_with_spike}.

Compared to the top left panel of Fig.~\ref{fig:BDM_flux_millennium}, the presence of the DM spike profile leads to a $\sim 12\%$ increase of the time-averaged MW BDM value from 8.09 to 9.14 $\times 10^{-14}~{\rm keV}^{-1}\,{\rm cm}^{-2}\,{\rm s}^{-1}$. 
For the DBDM flux, it also increases from 2.42 to 2.97 $\times 10^{-13}~{\rm keV}^{-1}\,{\rm cm}^{-2}\,{\rm s}^{-1}$.
Although the DM spike leads to slightly larger temporal variation of the MW~BDM flux, the values remain lower than the DBDM flux all the time.

We also remark that if the SN locations follow the H$_2$ gas distribution discussed in Appendix~\ref{app:H2_density}, 
then the effect of DM spike becomes negligible due to the fact that there should be almost no SN that occurs close to the GC (see Fig.~\ref{fig:normalized_density_profile}).

\section{Results with energy-dependent $\sigma_{\chi\nu}$}\label{app:energy_dependent_sigxv}

\begin{figure}
\begin{centering}
\includegraphics[width=0.9\columnwidth]{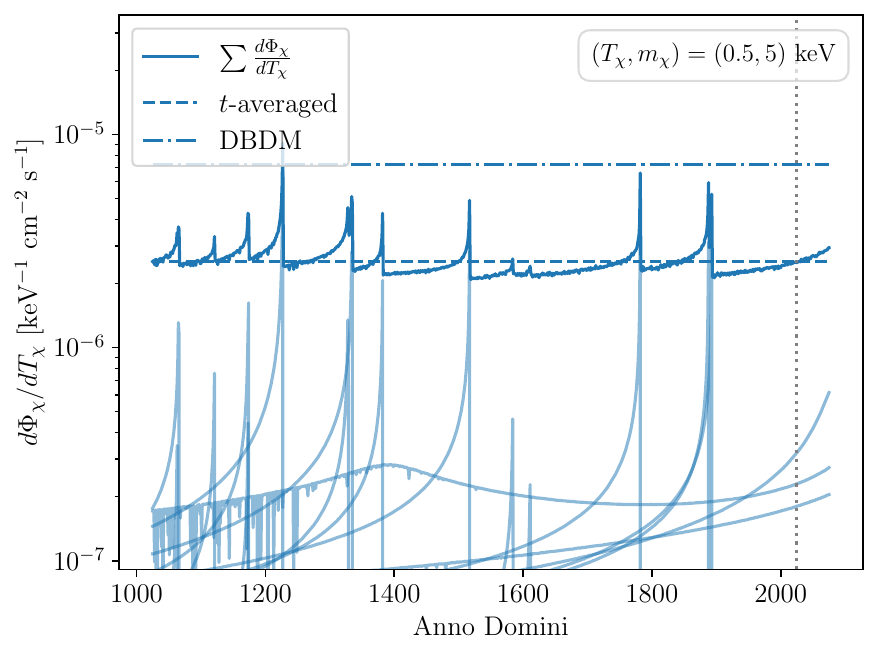}
\end{centering}
\caption{\label{fig:nonrelativistic_flux_energy_dependent}Evolution of BDM fluxes from MW SNe over a time span of one millennium with $d\sigma_{\chi\nu}/d\Omega$ derived from $L_\mu-L_\tau$. The SNe sample is the same as shown in Fig.~\ref{fig:sne_scattering}. Legends are identical to Fig.~\ref{fig:BDM_flux_millennium}.}
\end{figure}

The results presented in the main text are based on $\sigma_{\chi\nu}$ that is energy-independent. 
In this appendix, we adopt the
$L_\mu-L_\tau$ model \cite{Chang:2018rso,Croon:2020lrf,Escudero:2019gzq,Foldenauer:2018zrz} 
to explore how our results depend on this assumption.
The relevant terms in the Lagrangian to the discussion are
\begin{equation}
    \mathcal{L}\supset g_V V_\mu \bar{\nu}\gamma^\mu \nu + g_\chi V_\mu \bar{\chi}\gamma^\mu \chi - \frac{1}{2}m_V^2 V_\mu V^\mu - m_\chi \bar{\chi}\chi
\end{equation}
where $\nu$ is the neutrino field, $\chi$ is the fermionic DM field with mass $m_\chi$, $V_\mu$ is the vector mediator with mass $m_V$ and $g_{V,\chi}$ is the coupling constants.
The associated DM-$\nu$ scattering amplitude square is
\begin{equation}\label{eq:amplitude}
    |\mathcal{M}|^{2} =\frac{g_V^2 g_\chi^2}{(t-m_{V}^{2})^2}(s^{2}+u^{2} +4tm_\chi^2 -2m_\chi^4), 
\end{equation}
where $s,t,$ and $u$ are the Mandelstam variables.
The differential DM-$\nu$ cross section takes the same form as Eq.~\eqref{eq:diff_sigxv} but with the replacement \cite{Lin:2023nsm}
\begin{equation}
    \frac{d\sigma_{\chi\nu}}{d\cos\psi}=\frac{1}{32\pi}\sqrt{\frac{1}{m_\chi^3}\left(\frac{1}{m_\chi}+\frac{2}{T_\chi}\right)}|\mathcal{M}|^2.
\end{equation}
To proceed, we assume $(g_V,g_\chi)=(10^{-10},10^{-3})$ and $m_V=m_\chi/3$, which 
satisfies the current constraints \cite{Croon:2020lrf,Lin:2023nsm}. 

\begin{figure*}
\begin{centering}
\includegraphics[width=0.99\columnwidth]{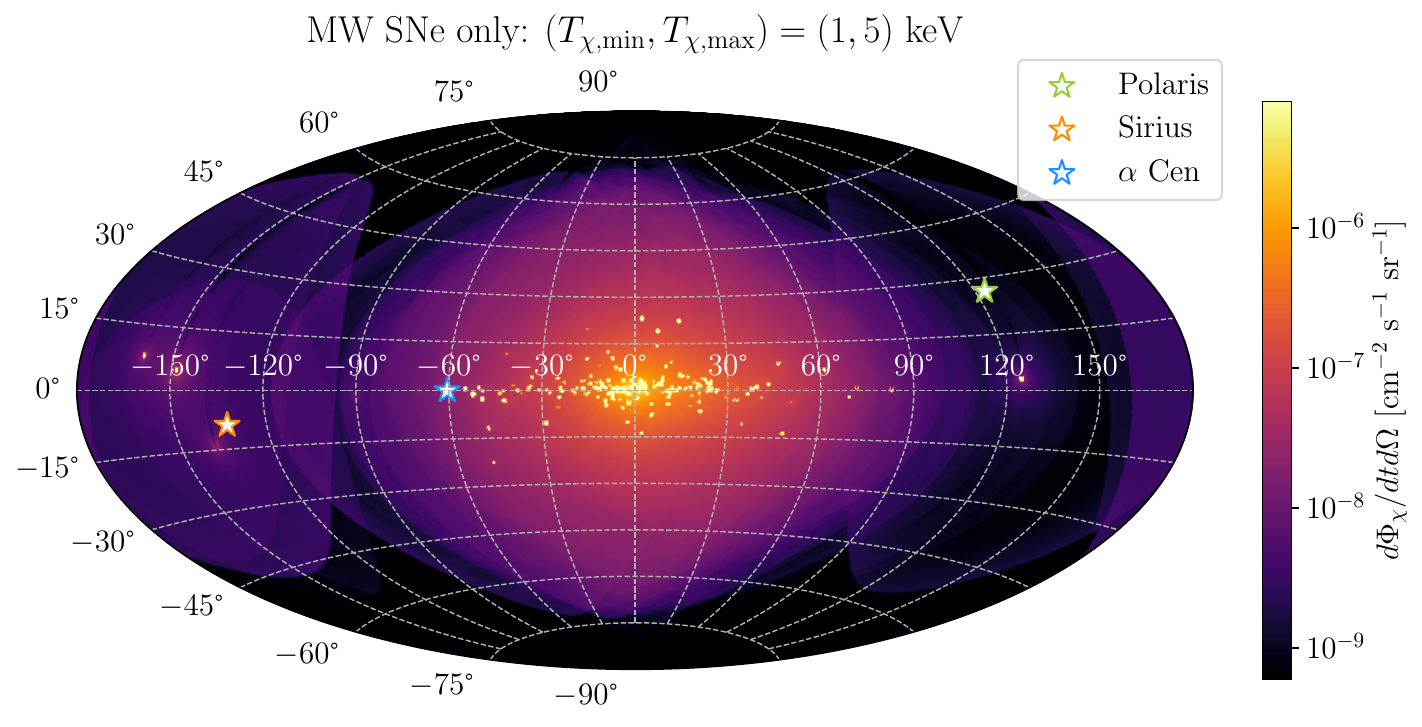}
\includegraphics[width=0.99\columnwidth]{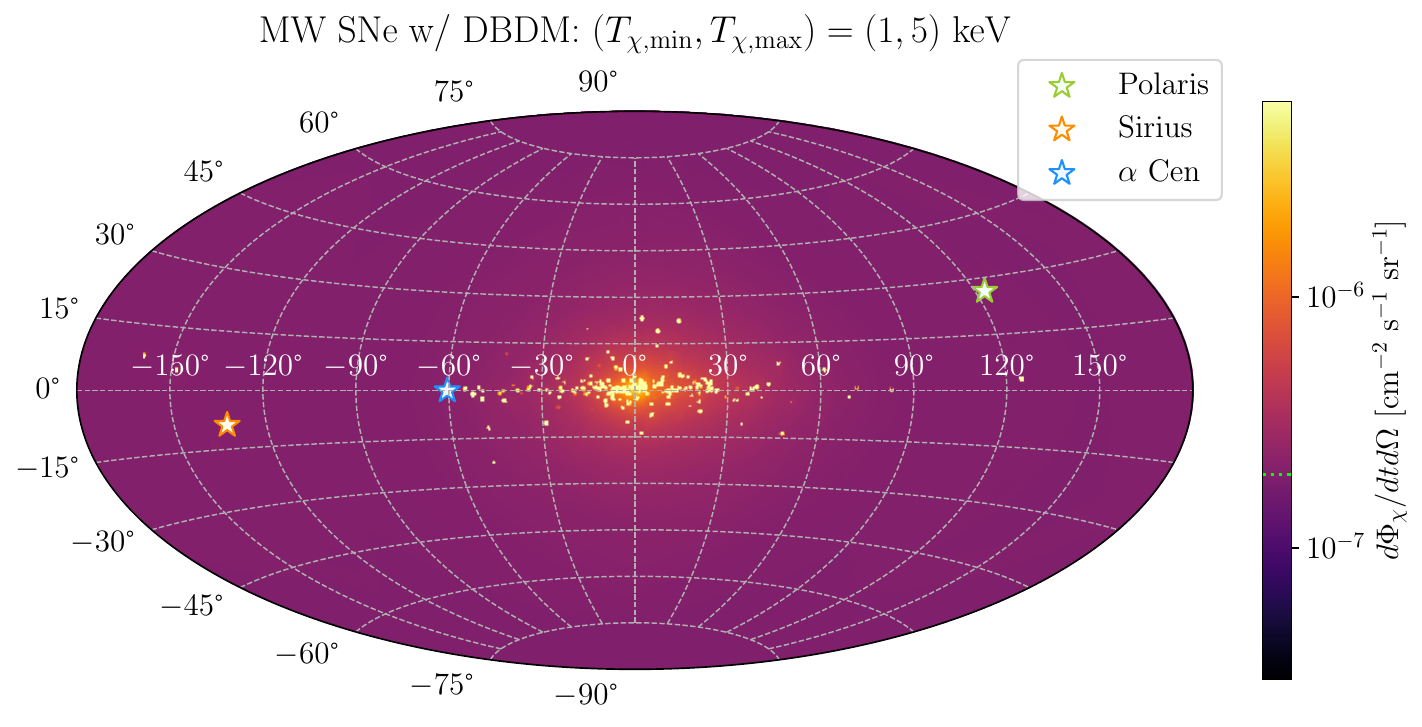}
\end{centering}
\caption{\label{fig:allsky_energy_dependent}
$T_\chi$-integrated BDM flux across the celestial sphere with $d\sigma_{\chi\nu}/d\Omega$ derived from the $L_\mu-L_\tau$ model. Left and right panels are without and with DBDM contribution. Its value is $1.945\times 10 ^{-7}$ cm$^{-2}$ s$^{-1}$ sr$^{-1}$ and is indicated by the short green dotted lines in the color bar.}  
\end{figure*}
The temporal evolution of the MW SN$\nu$ BDM flux over the past millennium for this model as well as the corresponding DBDM flux are shown in Fig.~\ref{fig:nonrelativistic_flux_energy_dependent}. 
Once again, despite the larger temporal variation of the MW BDM flux, it remains below the DBDM flux at all times.
We note that here the absolute value of the fluxes shown in the plot should not be directly compared to cases with energy-independent $\sigma_{\chi\nu}$. 

Additionally, we show in Fig.~\ref{fig:allsky_energy_dependent} the $T_\chi$-integrated BDM flux per steradian over the celestial sphere. 
The left and right panels show the angular distribution of MW~BDM only and the distribution taking into account the isotropic DBDM component, respectively. 
Although taking an energy-dependent model leads to a different BDM angular distribution at the boost location from what is shown in Fig.~\ref{fig:ang_dist}, the resulting sky map of the $T_\chi$-integrated BDM flux is qualitatively very similar to that obtained in the main text and is buried under the DBDM component ( see the top panels of Fig.~\ref{fig:allsky_map}).

 % Blue color ends here

\section{Derivation of approximated $t_{\rm van}$}\label{app:t_van_approx}

In this appendix we provide detailed derivation for the approximated formula of $t_{\rm van}$ [Eq.~\eqref{eq:t_van_approx} in the main text] in the relativistic ($m_\chi/T_\chi\ll 1$) and the nonrelativistic ($m_\chi/T_\chi\gg 1$) limits.

\subsection{$m_\chi/T_\chi\ll 1$}
Setting $x=m_\chi/T_\chi\ll 1$ and $c=1$ for simplicity and keeping 
all terms up to the leading order of $\mathcal{O}(x)$, 
one can approximate the BDM velocity $v_\chi\approx 1$ and the maximal scattering angle 
$$
    \psi_{\rm max}\approx\cos^{-1}\left(\frac{1}{\sqrt{1+2x}}\right)\approx \sqrt{2x}.
$$
Rewriting
Eq.~\eqref{eq:theta*} in the main text 
as
$$
    \tan\theta^* = \frac{v_\chi-\cos\psi_{\rm max}}{\sin\psi_{\rm max}}
$$
and using
$$
\cos\psi_{\rm max} \approx 1-x, 
$$
one gets 
\begin{equation}
    \tan\theta^* \approx \frac{x}{\sqrt{2x}}=\sqrt{\frac{x}{2}}\quad \to \quad \theta^*\approx \sqrt{\frac{x}{2}} 
\end{equation}
for $x\ll 1$. 
Rewriting Eq.~\eqref{eq:tvan} in the main text as 
\begin{equation}\label{eq:tvan_ll_app}
    t_{{\rm van}}  =R_{s}\frac{\sin\theta^{*}}{\sin\psi_{{\rm max}}}+\frac{R_{s}}{v_{\chi}}\frac{\sin(\psi_{{\rm max}}-\theta^{*})}{\sin\psi_{{\rm max}}}-t_{\nu} 
\end{equation}
and approximating the first term by 
$$
\frac{\sin{\theta^*}}{\sin\psi_{\rm max}}=\frac{\sin{\sqrt{x/2}}}{\sin\sqrt{2x}}
\approx \frac{1}{2}+\frac{x}{8}
$$
as well as the second term by 
$$
\frac{\sin(\psi_{{\rm max}}-\theta^{*})}{\sin\psi_{{\rm max}}} \approx \frac{1}{2}+\frac{x}{8}, 
$$
Eq.~\eqref{eq:tvan_ll_app} reduces to 
\begin{equation}
    t_{\rm van} \approx R_s\left(
    \frac{1}{2}+\frac{x}{8}+\frac{1}{2}+\frac{x}{8}-1
    \right) = \frac{x R_s}{4}.
\end{equation}
Restoring $c$ leads to Eq.~\eqref{eq:tvan} for $x\ll 1$ in the main text.

\subsection{$m_\chi/T_\chi\gg 1$}

For $x\gg 1$, we expand $v_\chi$ and $\psi_{\rm max}$ and keep all terms up to $\mathcal{O}(1/\sqrt{x})$.
This gives  
$v_\chi\approx \sqrt{2/x}$ and
$$
    \psi_{\rm max}\approx\cos^{-1}\left(\frac{1}{\sqrt{2x}}\right)\approx \frac{\pi}{2}-\frac{1}{\sqrt{2x}}.
$$
Taking these expressions, Eq.~\eqref{eq:theta*} in the main text becomes
\begin{equation}
    \tan\theta^* = \frac{v_\chi-\cos\psi_{\rm max}}{\sin\psi_{\rm max}}\approx \frac{1}{\sqrt{2x}} \quad \to \quad \theta^*\approx \frac{1}{\sqrt{2x}},
\end{equation}
such that
\begin{alignat}{1}
t_{{\rm van}} & =R_{s}\frac{\sin\theta^{*}}{\sin\psi_{{\rm max}}}+\frac{R_{s}}{v_{\chi}}\frac{\sin(\psi_{{\rm max}}-\theta^{*})}{\sin\psi_{{\rm max}}}-t_{\nu}\nonumber \\
 & \approx R_{s}\left(1+\frac{1}{v_{\chi}}-1\right) = R_{s}\sqrt{\frac{x}{2}}.\label{eq:tvan_gg_app}
\end{alignat}
Once again, restoring $c$ recovers  Eq.~\eqref{eq:tvan} in the main text for $x\gg1$. 

We note that the peak time defined in Eq.~\eqref{eq:tp} in the main text is approximately
\begin{equation}
    t_p \approx R_s \left( \sqrt{\frac{x}{2}} - 1\right) \approx R_s \sqrt{\frac{x}{2}}, 
\end{equation}
which is identical to  Eq.~\eqref{eq:tvan_gg_app}.
Thus, $t_p$ and $t_{\rm van}$ coincide under $x\gg1$.

%\bibliography{main}

\end{document}